\documentclass[a4paper,11pt]{article}
\usepackage{graphicx}
\usepackage[space]{grffile}
\usepackage{latexsym}
\usepackage{textcomp}
\usepackage{longtable}
\usepackage{tabulary}
\usepackage{booktabs,array,multirow}
\usepackage{amsfonts,amsmath,amssymb}
\usepackage{natbib}
\usepackage{url}
\usepackage{hyperref}
\hypersetup{colorlinks=false,pdfborder={0 0 0}}
\usepackage{etoolbox}
\makeatletter
\usepackage{authblk}
\usepackage[utf8]{inputenc}
\usepackage[english]{babel}
\usepackage[top=1in, bottom=1in, left=1in, right=1in]{geometry}
\title{Annual cycle and longitudinal structure of tropical eddy and mean momentum fluxes}
\date{}
\author[1,2]{Abu Bakar Siddiqui Thakur \thanks{Corresponding author: A B S Thakur, thakur.abubakar@gmail.com}
}
\author[1,2]{Jai Sukhatme}
\affil[1]{Centre for Atmospheric and Oceanic Sciences, Indian Institute of Science, Bangalore 560012, India}
\affil[2]{Divecha Centre for Climate Change, Indian Institute of Science, Bangalore 560012, India}

\begin{document}

\maketitle

\begin{abstract}

The longitudinal structure and annual cycle of mean meridional and eddy momentum fluxes in the tropical upper troposphere are studied. In zonal mean, these two terms oppose each other and peak during the Indian summer monsoon. This zonal mean character arises from a rich longitudinal structure that is revealed by splitting the globe into three zones, namely, the Asia-West Pacific (AWP), central Pacific-West Atlantic (CP-WA) and African sectors. The mean convergence term is cohesive across all three regions, has a single peak in the boreal summer and always acts to decelerate the zonal flow. A Helmholtz decomposition shows that the advection of absolute vorticity by the divergent meridional wind in localized cross-equatorial cells is responsible for the coherent nature of the mean convergence across all sectors. \\

\noindent On the other hand, the eddy convergence goes from being small and seasonally invariant in the African region to one with large seasonal maxima (minima) in AWP (CP-WA) sector that accelerate (decelerate) the zonal mean flow. The disparate nature the eddy flux in the AWP and CP-WA regions is in the winter season and is due to the tropical and extratropical origin of waves, respectively. In summer, the AWP region accounts for almost all of the eddy flux convergence. In fact, the leading role of the rotational zonal - divergent meridional component in the zonal mean eddy flux does not hold in individual sectors. Finally, through the year, the CP-WA region is where the local overturning cell is strongly influenced by eddy activity. \\


\end{abstract}

\section{Introduction}

The Hadley Cell (HC) is the dominant feature of the zonally averaged tropical tropospheric circulation.
Towards the later half of the twentieth century, theoretical efforts aimed at understanding the HC emphasized axisymmetric dynamics with rising motion at the equator forced by latent heating of cumulus convection and subsidence in the subtropics \citep{schneider1977axially,schn1977}. 
In this context, \citet{held1980nonlinear} arrived at simple analytical relations between parameters predicting the latitudinal extent and strength of the tropical overturning motion. 
Extensions to a moist setting \citep{satoh} as well as the time dependent nature of the flow \citep{fangtung} were also explored in this axis-symmetric scenario. 
The cross-equatorial nature of the cell and an abrupt threshold behaviour of the angular momentum conserving zonal mean circulation in response to off-equatorial heating was elucidated by \citet{lindzen1988hadley} and \citet{plumbhou}, respectively. 
The nature of this axis-symmetric cell with subtropical heating has also been studied in a moist framework with an eye towards explaining the rather sudden onset of the south-Asian monsoon flow \citep{eman95,booseman}.

\noindent Though the possible importance of non-axis-symmetric dynamics on the tropical circulation has been known for a long time \citep{kuo, schneider1984response}, the role of midlatitude eddies in influencing the HC has been brought to the fore in recent years \citep{bordoni-schneider2010,Singh}. Indeed, three dimensional primitive equation based simulations and general circulation model experiments in a host of climate regimes have seen to yield results that differ from axis-symmetric predictions \citep{becker1997,kimlee,walker2006eddy,frierson,korty}.
In particular, \citet{walker2006eddy} showed that the HC falls in regimes intermediate to angular momentum conserving (flow responds directly to changes in thermal driving) and eddy-driven limits. 
This has also been found to be applicable in the seasonal cycle of the overturning motions \citep{schneider2008eddy, bordoni2008monsoons}. In fact, recent results highlight the role of eddies in allowing the HC extend down to the surface from the upper troposphere \citep{Davis}. 
Further, studies using reanalysis data have provided evidence of extratropical eddy influences on the tropics on interannual timescales \citep{caballero2007role,caballero2009impact}. 

\noindent In all, given their prominent role, it is important to understand the nature and transport of momentum by the eddies in the deep tropics.
In this context, an examination of the upper tropospheric annually averaged zonal momentum budget showed that eddies induce a westerly acceleration in a zonal mean sense \citep{lee1999climatological}. 
This is overwhelmed by an easterly acceleration due to the mean flow advection of zonal momentum by the seasonally reversing HC and causes the equatorial upper tropospheric flow to be easterly in nature.
Further, the westerly acceleration in the equatorial upper troposphere was shown to be due to the convergence of eddy momentum flux by the thermally forced climatological Rossby gyres present throughout the year \citep{dima2005tropical}. More detailed analysis revealed that the eddy momentum flux convergence in the deep tropics is dominated by correlations between the rotational zonal and divergent meridional wind \citep{zurita2019role}.
These tropical stationary eddies, forced by longitudinal thermal contrasts \citep{wang1999seasonal}, have an inherent zonal dependence which is lost in averaging.
But, even in the context of the zonal mean picture, there are a number of basic questions that arise here. 
For example, does the eddy momentum flux have a similar character through the year? 
Does the flux convergence always accelerate the zonal mean flow?
Given that the extratropical waves can reach the tropics \citep{hoskins1993rossby}, is there a competition between eddies of differing origin?
In other words, are there periods in the year where the eddies of a midlatitude origin ``win" and the eddy fluxes actually act to decelerate the mean zonal flow that is part and parcel of the HC?

\noindent Along with the recognition of the role of eddies, it is also important to note that the tropical overturning circulation is longitudinally localized.
These localized circulations result from the longitudinal heterogeneity in the distribution of thermal forcing associated with the world's monsoons.
Although, the pioneering study of \cite{gill1980} formulated the steady linear response of the atmosphere to a localized heat source, interest in the zonal structure of the response has re-awakened recently and
most studies, in this regard, have focused on the interannual variability of the resulting longitudinally limited overturning motions \citep{zhang2013interannual,nguyen2018variability,sun2019regional}.
\citet{hoskins2019detailed} presented a detailed analysis of the Northern Hemisphere (NH) summer HC and showed that the Asian monsoon circulation is the dominant contributor to the zonally averaged circulation during the boreal summer. 
Further, the rapid transition of southern cell during the Asian monsoon from an eddy-driven regime to a thermally-driven one \citep{bordoni2008monsoons} was seen to be not so abrupt when the forcing itself was zonally asymmetric \citep{ZhaiBoos}. 
Indeed, regime transitions of the HC have been examined in the presence of inhomogeneous boundary forcing \citep{shaw2014role,geen}. 
More insight into the localised nature of the HC has been possible through a coupled Eulerian-Lagrangian analysis that suggests a "conveyor belt" type circulation with large-scale ascent in the Indo-Pacific region and subsequent descent over the Americas \citep{raiter2020tropical}. 

\noindent Thus, bringing the two threads together, the issue at hand is to understand the temporal variation and longitudinal nature of eddy fluxes in the deep tropics and subsequently put them in perspective with the local overturning circulations.
In essence, the questions we posed above need to be taken a step further and our aim in this study is to first of all study the annual cycle of momentum fluxes and then probe these longitudinal variations in eddy and mean fluxes in the tropical upper troposphere. 
The outline of the paper is as follows: In Section 2 we describe the data used and the methods employed to examine the regional tropical HC and momentum fluxes. Section 3 deals with the annual cycle of the momentum fluxes. Section 4 presents these fluxes and the regional overturning cells during the NH winter, summer and spring equinoctial period. Section 5 contains a summary of results and their discussion. 

\section{Data and Methods}

\noindent The data used in this study comprises of four-times daily horizontal wind at a resolution of $2.5^{\circ}$ across 17 pressure levels from ERA-Interim \citep{dee2011era} for a 40-year period from 1979-2018. We also make use of monthly mean GPCP Precipitation data \citep{adler2003version} with the same resolution and over the same period provided by NOAA (\url{https://psl.noaa.gov/}).

\noindent The zonally averaged zonal momentum equation reads \citep{dima2005tropical,kraucunas2005equatorial},
\begin{equation}\label{eq:1}
    \frac{\partial [u]}{\partial t} = 
    [v] \left(f - \frac{1}{\cos\phi}\frac{\partial [u] \cos\phi} {\partial y} \right) - \frac{1}{\cos^2 \phi}\frac{\partial [u^*v^*]\cos^2 \phi}{\partial y} - [\omega]\frac{\partial [u]}{\partial p} - \frac{\partial [u^* \omega^*]}{\partial p} + [\overline{X}]
\end{equation}

\noindent The notation above is standard. Specifically, square braces denote a zonal mean and asterisks denote a deviation from this mean. The first term on the right is the mean meridional momentum flux convergence, the second term is the meridional eddy momentum flux convergence. The third and fourth terms are the mean vertical advection and vertical eddy flux convergence. Consistent with previous studies, we find these to be small in comparison to the first two terms on the RHS. 
The last term is a residual and accounts for all sub grid-scale processes. The first part of this study focuses on the Day of Year variation of the first two terms on the RHS of Equation \ref{eq:1}. In particular, daily estimates of each variable are calculated and then these are averaged over the respective days through the 40 years on record. 

\noindent Utilising the flux decomposition of \cite{lee1999climatological}, the eddy and mean flow convergence terms of Equation \ref{eq:1} can broken down into five components. Specifically,  

\begin{equation}\label{eq:2}
    - \frac{1}{\cos^2\phi}\frac{\partial{[\overline{u^*v^*}]\cos^2\phi}}{\partial y} = - \frac{1}{\cos^2\phi} \frac{\partial{[\bar{u^*}\bar{v^*}]\cos^2\phi}}{\partial y} - \frac{1}{\cos^2\phi} \frac{\partial{[\overline{u^{*'}v^{*'}}]\cos^2\phi}}{\partial y}
\end{equation}

\begin{equation}\label{eq:3}
    \overline{[v] \left(f - \frac{1}{\cos \phi} \frac{\partial {[u] \cos \phi}}{\partial y} \right) } = f [\overline{v}] - [\overline{v}] \frac{1}{\cos \phi} \frac{\partial {[\overline{u}] \cos \phi}}{\partial y} - \overline{[v'] \frac{1}{\cos {\phi}} \frac{\partial {[u'] \cos \phi}}{\partial y}}
\end{equation}
\noindent The two terms on the RHS of Equation \ref{eq:2} are momentum convergences associated with the stationary eddies (SE) and transient eddies (TE). The three terms on the RHS of Equation \ref{eq:3} are due to the Coriolis force (CF), momentum convergences linked to mean meridional circulation (MMC) and transient meridional circulation (TMC). Apart from the notation presented before, overbars indicate a temporal mean and primes indicate a deviation from this mean. 

\noindent For part of our analysis, we make use of a rotational-divergent partition. Specifically, treating the horizontal wind field on each pressure level as a two-dimensional vector field, we split it into rotational and divergent components via a Helmholtz decomposition. The horizontal velocity field is expressed as, 
\begin{equation}\label{eq:4}
    \Vec{v} = \nabla \chi - \Vec{k} \times \nabla \psi.
\end{equation}
The first term on the RHS describes the irrotational component while the second term on the RHS describes the non-divergent component of the velocity field. The rotational and divergent components will be denoted by the subscripts $r$ and $d$ respectively.

\noindent Seasonal averages of these terms are computed by time averaging the respective quantities over the duration of a season for every year on record and then presented as a 40-year mean. The major seasons considered in this study are NH winter, spring and summer. To obtain results that are consistent with and comparable against previous studies, the seasonal estimates are computed by temporal averaging over the dates mentioned in Table 1 of \citet{dima2005tropical}. Seasonal estimates of winter, summer and spring are presented.

\section{Annual Cycle of Momentum Fluxes}

\begin{figure}[ht]
    \centering
    
    \includegraphics[width=0.8\linewidth]{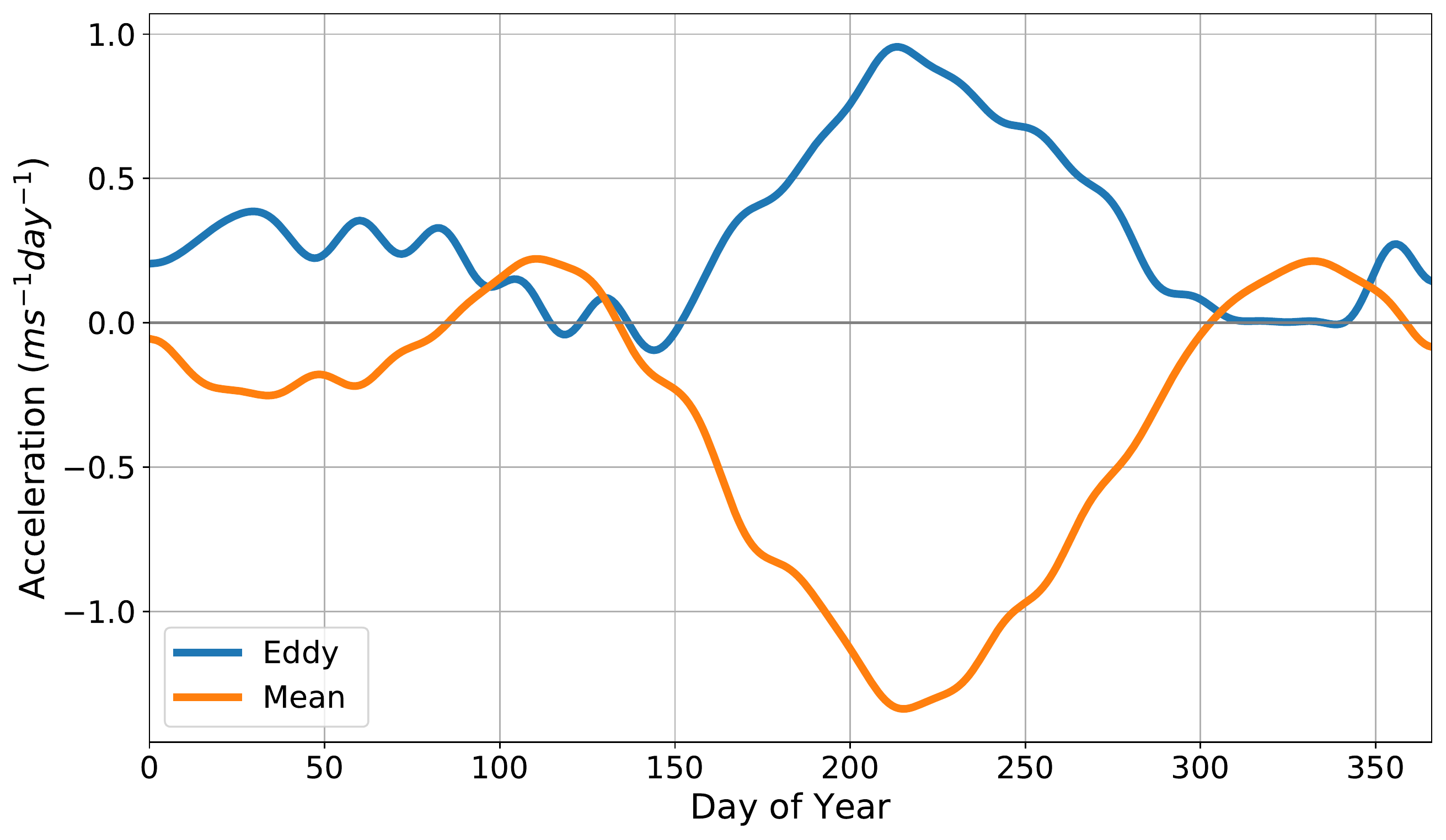}
    \caption{Climatologial Day of Year variation of the zonally averaged mean meridional momentum flux convergence (orange) and eddy momentum flux convergence (blue), as Equation \protect \ref{eq:1}, averaged over 150-300 hPa, about $\pm 5^{\circ}$ of the equator. A 20-day low-pass filter has been applied prior to presentation.}   
    \label{fig:eddyVsMean}
\end{figure}

\noindent Figure \ref{fig:eddyVsMean} describes the annual cycle of the zonal mean eddy momentum flux convergence and the mean meridional momentum flux convergence, i.e., the two major terms in Equation \ref{eq:1} for the zonally averaged momentum budget of the tropical upper troposphere. In an annual mean, the mean meridional term exceeds the eddy convergence and leads to the tropical upper tropospheric easterlies \citep{lee1999climatological}.
Compared to the rest of the year, both the mean and eddy momentum convergence terms are particularly enhanced during the Asian summer monsoon season (June through September) and are close to zero during the equinoctial periods. There is a strong sense of anti-correlation (correlation coefficient of -0.93, Table \ref{table:correlation}) between the mean meridional and eddy momentum flux convergence throughout the year. This has been noted previously on seasonal timescales \citep{dima2005tropical,kelly2011zonal}. 
In fact, while emphasising on the roles of planetary and sub-planetary scale eddies in the winter-summer transition of the NH circulation, a similar balance between mean meridional convergence and planetary scale eddy flux convergence was noted \citep{shaw2014role}. 

\noindent The dynamical phenomena behind this compensation in the upper tropospheric zonally averaged zonal mean momentum budget have also been explored previously. \cite{dima2005tropical} observed that there is a coincidence between the zonally averaged eddy flux and mean meridional winds in the upper troposphere along with a negative correlation between the climatological $[v]$ and $-\partial_y{[u]}$. They argue that this tendency for opposition between the two convergence terms occurs because of the preference of the zonal mean tropical rain belts and eddy forcing to occur in the same latitude band. The zonally averaged heating generates the meridional overturning circulation and a resultant upper tropospheric mass flux divergence, while the eddy forcing results in a momentum flux convergence into the source region.
In fact, stationary Rossby wave energy can propagate meridionally through easterlies in the presence of a meridional background flow \citep{watterson1987effect, li2015interhemispheric}.
This line of reasoning is supported by idealised modelling efforts \citep{kraucunas2007tropical}. 
Further, the imbalance created by any eddy flux convergence is nullified by a mean flow adjustment to maintain thermal wind balance \citep{kelly2011zonal}.

\begin{figure}[ht]
    \centering
    
    \includegraphics[width=0.8\linewidth]{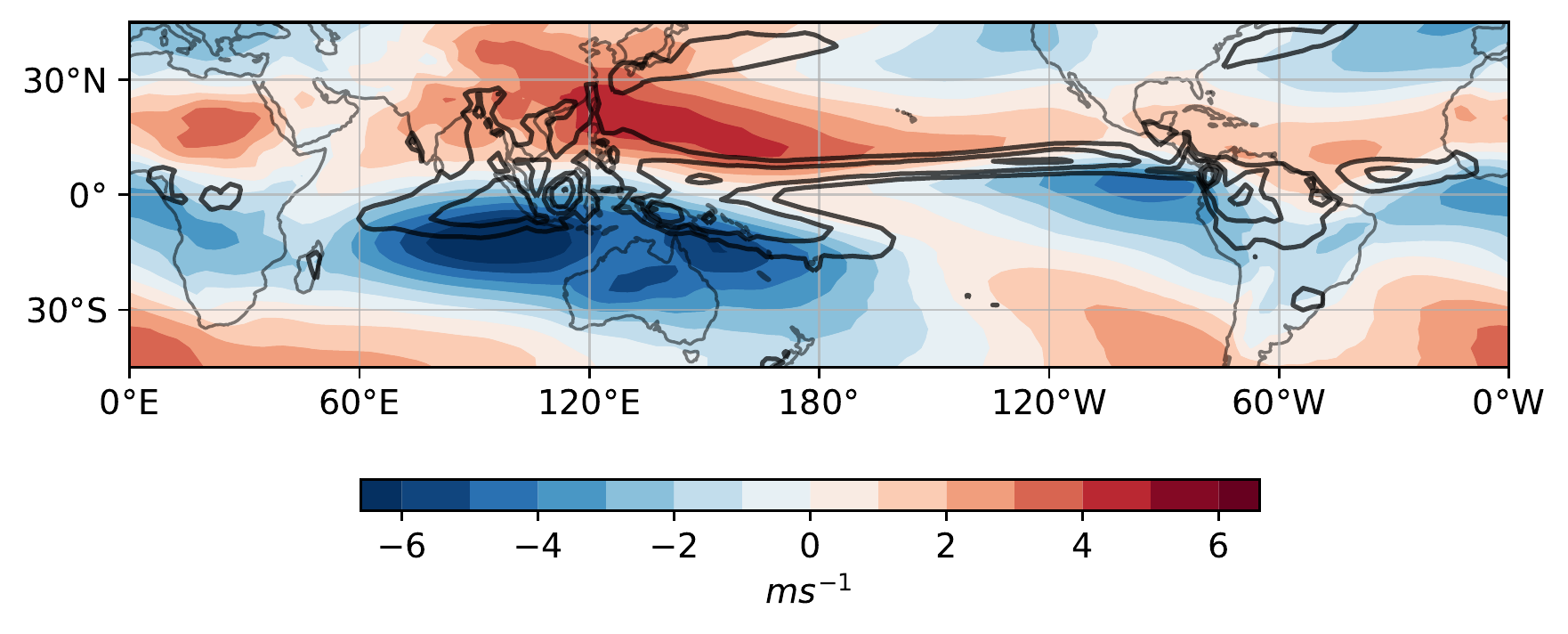}
        \caption{Spatial map of the annual mean difference between upper (150hPa) and lower level (925 hPa) divergent meridional winds  (colors), along with time mean precipitation (contours). The contours of precipitation are at 5., 7., 9. and 11. mm/day. 
        }
    \label{fig:diffVd}
\end{figure}

\noindent Recent work highlights that the seasonally varying zonally averaged mean meridional circulation is actually composed of longitudinally limited overturning circulations \citep{hoskins2019detailed}. Following the argument that it is the divergent motions which contribute to the north-south meridional overturning circulations \citep{zhang2013interannual, schwendike2014local}, we construct a spatial map of the difference between the upper and lower tropospheric divergent meridional wind as an annual mean overlaid with contours of annual mean precipitation.
Such a difference strengthens the upper level divergent motion forced by the monsoonal heating.
Figure \ref{fig:diffVd} shows that, the tropical overturning activity can be partitioned into three sectors. They are Asia-West Pacific (40E - 150W; abbreviated as AWP), Central Pacific - West Atlantic (150W - 30W; CP-WA) and African (30W - 40E) regions.
These three sectors have been motivated by the regional monsoonal circulations that exist within them.
Of course, the exact choice of longitudinal extents for each regional sector is subjective, the results are roughly insensitive to such alternatives. 

\begin{figure}[ht]
    \centering
    
    \includegraphics[width=0.8\linewidth]{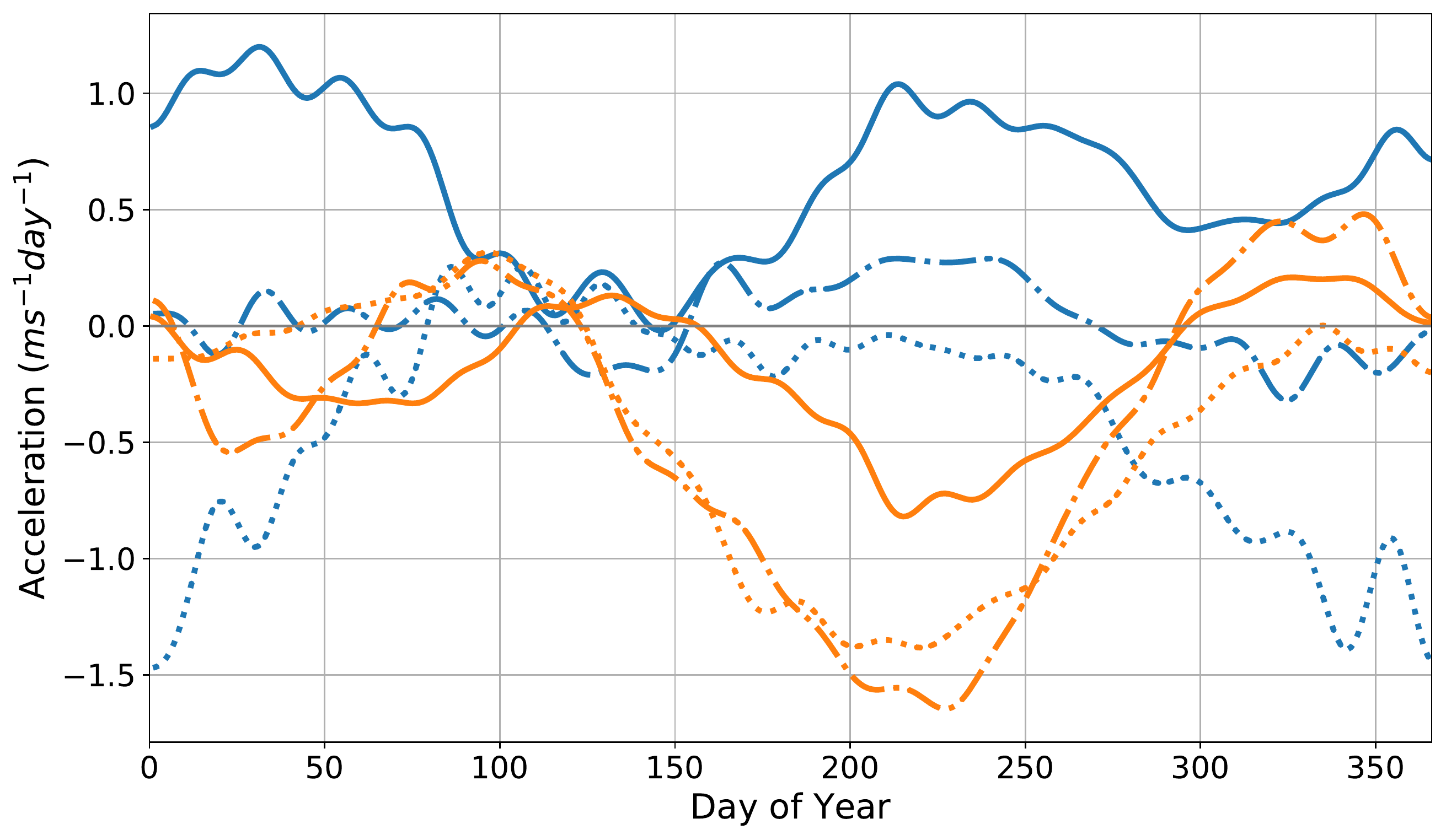}
    \caption{Same as Figure \protect \ref{fig:eddyVsMean} except that the zonal mean and eddy computations are over respective longitudinal bands mentioned in reference to Figure \protect \ref{fig:diffVd}. The thick solid lines represent the momentum flux convergence computed and averaged over AWP sector, the thinner dotted lines are for those over CP-WA and the broken dash-dot lines are for the African regions respectively.}   
    \label{fig:eddyVsMean3zones}
\end{figure}

\noindent The solid lines in Figure \ref{fig:eddyVsMean3zones} show mean and eddy momentum flux convergences as in Figure \ref{fig:eddyVsMean}, but when computed only over the AWP sector. 
This region, covering the Asian summer monsoon and Australian monsoon sectors, is larger and associated with stronger overturning motions than the other two longitudinal zones mentioned with reference to Figure \ref{fig:diffVd} \citep{hoskins2019detailed}. On comparison with the quantities computed over the global domain in Figure \ref{fig:eddyVsMean}, the two mean convergence terms agree quite well with each other (thick orange curves in Figures \ref{fig:eddyVsMean} and \ref{fig:eddyVsMean3zones}; correlation coefficient of 0.88 as in Table \ref{table:correlation}), but the eddy convergences are quite different during a large part of the year (thick blue lines; 0.59). This disagreement between the two eddy momentum convergence estimates begins around the decay of the Indian monsoon (approximately Day 220) and persists into spring season of the following year. Like the global domain, the mean and the eddy momentum flux convergences computed over this region also exhibit a weaker anti-correlation of -0.56. The eddy momentum flux convergence in this region is almost always positive and shows two peaks, one in the winter and one in the summer season. The mean meridional momentum flux tries to retard the zonal mean flow and peaks in the summer season.

\begin{table}[ht]
\centering
\begin{tabular}{*4c}
    \multirow{2}{*}{Region}&\multirow{2}{*}{EMFC vs MMFC} & \multicolumn{2}{c}{Global vs regional} \\
    \multicolumn{2}{c}{}&  EMFC &    MMFC\\
    \hline
    Global  &  -0.93  &    -     &    -  \\
    AWP     &  -0.56  &   0.59   &  0.88 \\
    CP-WA   &  -0.27  &   0.30   &  0.93 \\
    Africa  &  -0.77  &   0.84   &  0.96 \\
\end{tabular}
\caption{Pearson's correlation coefficients computed (first column) between eddy momentum flux convergence (EMFC) and mean meridional momentum flux convergence (MMFC) over different regions (Figures \protect \ref{fig:eddyVsMean} and \protect \ref{fig:eddyVsMean3zones}), between global and regional (second column) EMFC  and (third column) MMFC.}
\label{table:correlation}
\end{table}

\noindent The set of dotted curves in Figure \ref{fig:eddyVsMean3zones} show the mean and eddy momentum flux convergence over CP-WA sector. This region is associated with the American monsoons and exhibits a divergence of eddy momentum flux over most of the year except during spring. The mean meridional flux is also mostly negative through the year, and in summer, is greater in magnitude than that computed over the global domain. Annual mean values of the eddy and mean meridional flux convergences over this region are $-0.4$ and $-0.44$ $ms^{-1}{day}^{-1}$, respectively. In contrast to the roles associated with the eddy and mean fluxes over global domain \citep{lee1999climatological}, both the flux convergence terms in the CP-WA region try to decelerate the flow over most of the year. From Table \ref{table:correlation}, the anti-correlation between the two fluxes here is weaker than in the AWP sector. Further, Table \ref{table:correlation} also shows that when compared against the global domain, the linear association between the two eddy (mean) convergences is weak (strong).

\noindent Fluxes computed over the African sector are shown in dash-dot lines in Figure \ref{fig:eddyVsMean3zones}. This region is associated with the African monsoons, and is smaller in longitudinal extent ($70^{\circ}$) compared to AWP ($170^{\circ}$) and CP-WA ($120^{\circ}$) but has overturning circulation of strength comparable to that of CP-WA region (illustrated by the strength of $V_d$ in Figure \ref{fig:diffVd}; consistent with Figure 5 of \citet{hoskins2019detailed}). The eddy momentum flux convergence in the African region is small ($\leq 0.3$ $ms^{-1}day^{-1}$) over the course of the annual cycle, although the mean meridional momentum flux convergence is very similar in form to that of the global domain (linear association of 0.96, Table \ref{table:correlation}) and has the largest peak of all the three regions.

\noindent In all, we see that the mean meridional flux is mostly negative in all regions and this is reflected in the global mean in Figure \ref{fig:eddyVsMean}. Interestingly, despite representing different longitudinal zones, the magnitude of this flux convergence in each zone peaks in the boreal summer. Further, the nature of the mean flux convergence is similar in all three regions throughout the year. Eddy fluxes on the other hand are much more diverse in character with minimal strength in the African sector through much of the year and a decelerating (accelerating) influence on the zonal flow in the CP-WA (AWP) region during winter. In addition, in AWP region, the eddy flux convergence peaks in both the boreal summer and winter. Whereas, in CP-WA, there is a single peak in the winter season. For a more fundamental understanding of these convergence terms, we now examine how the rotational and divergent components of the horizontal winds contribute to the total momentum fluxes. After which, to probe changing character of the eddy fluxes we consider the winter, summer and spring time balances separately.

\subsection{Rotational and Divergent Components}

\begin{figure}[ht]
    \centering
    
    \includegraphics[width=0.8\linewidth]{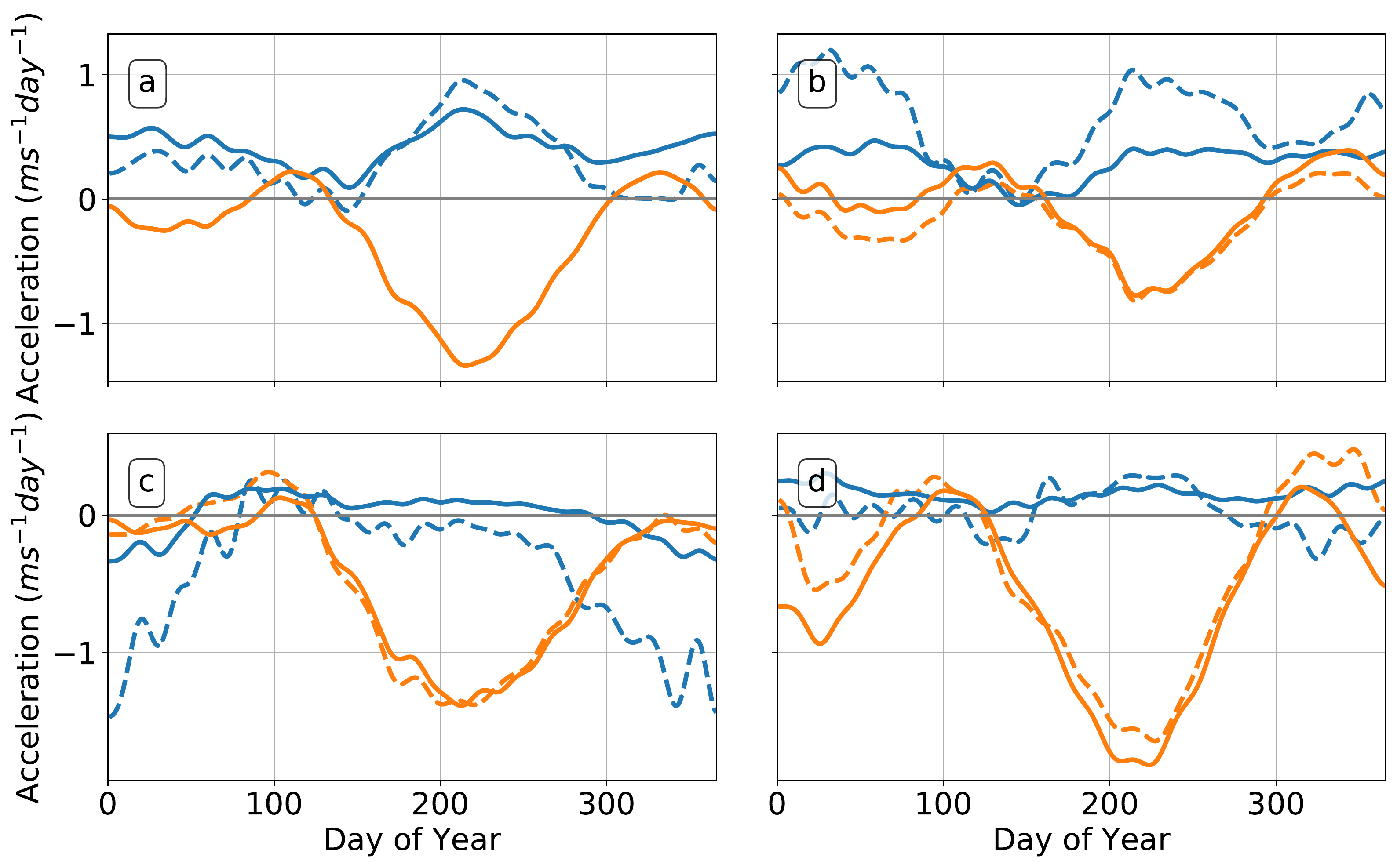}
    \caption{Same as Figures \protect \ref{fig:eddyVsMean} and \protect \ref{fig:eddyVsMean3zones} except that solid lines represent the $u_r v_d$ eddy and mean momentum flux convergences (Equation \protect \ref{eq:4}) computed over a) global longitudes, b) AWP, c) CP-WA and d) African sectors. The dashed lines in each panel represent the full eddy and mean flux convergences over each of these regions.}
    \label{fig:rotDiv}
\end{figure}

\noindent Figure \ref{fig:rotDiv} shows panels of the eddy and mean flux convergence computed using the rotational zonal wind and divergent meridional wind (solid lines) along with those computed using the total velocity field (dashed lines) over the individual sectors considered in this study. Convergence computed using other combinations of the rotational-divergent components of winds are not shown as the mean terms obtained from them are small compared to the ones computed using $u_r$ and $v_d$. For simplicity, we will use $uv$ and $u_r v_d$ to denote flux computed from the full velocity field and the $u_r$ - $v_d$ components respectively.

\noindent The most striking feature of Figure \ref{fig:rotDiv}a is that the $u_r v_d$ contribution and total mean convergence (solid and dashed orange curves) are identical. The reason behind this identity can be elucidated simply. On splitting Equation \ref{eq:4} into its zonal and meridional wind components, it can be seen that that $v_r$ and $u_d$ are zonal gradients of a stream function and velocity potential, respectively, and go to zero when integrated over all longitudes. Further, as seen in Figure \ref{fig:rotDiv}b,c and d, the dominance of the $u_r v_d$ contribution to the mean flux also holds individually in every region.

\begin{figure}[ht]
    \centering
    
    \includegraphics[width=0.8\linewidth]{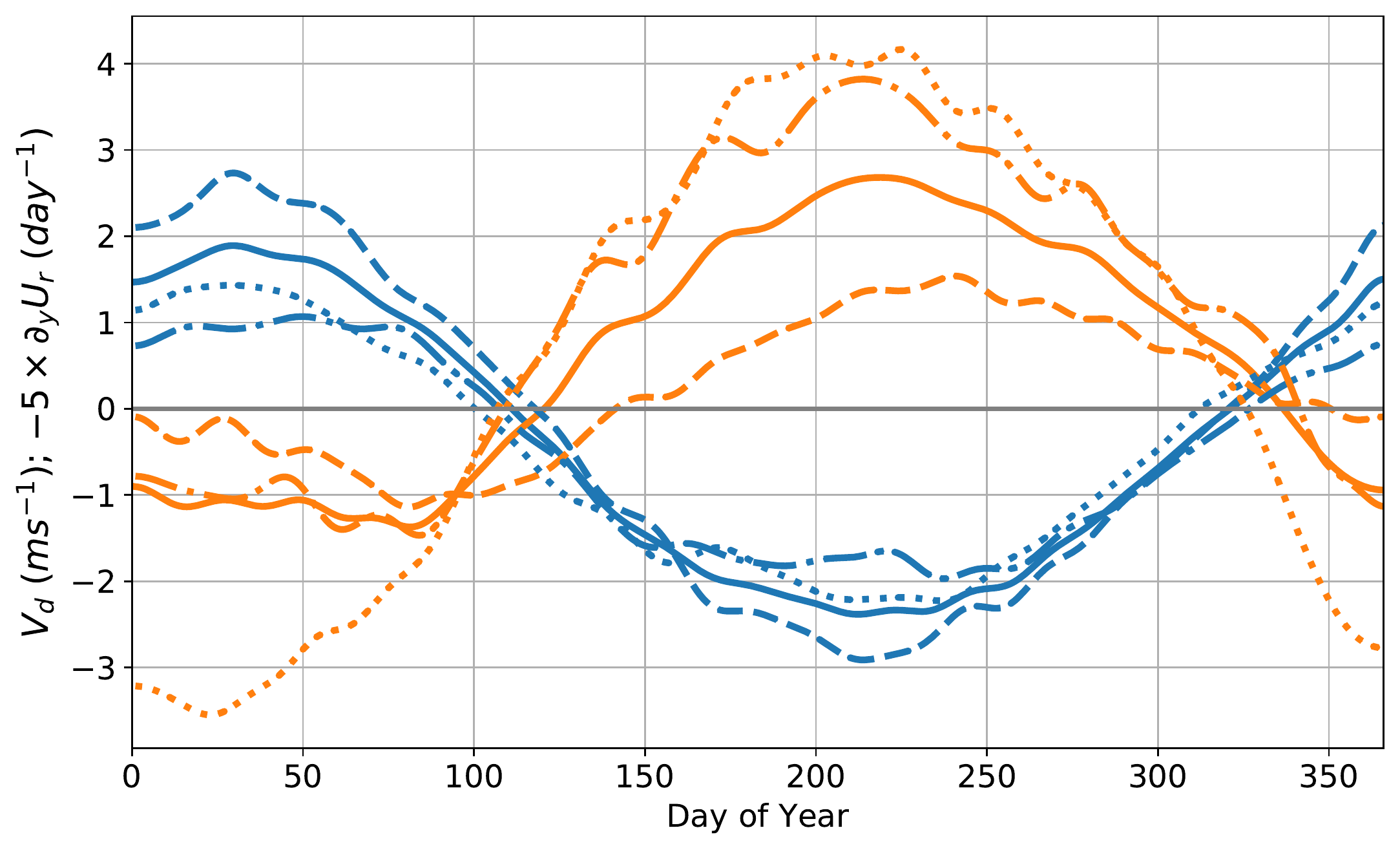}
    \caption{Climatological Day of Year variation of $v_d$ (blue) and $-5 \times \partial_y u_r$ (orange) zonally averaged over all longitudes (solid), AWP (dashed), CP-WA (dash-dotted) and Africa (dotted). As Figures \protect \ref{fig:eddyVsMean}, all quantities are averaged over 150-300 hPa, about $\pm 5^{\circ}$ and a 20-day low-pass filter has been applied prior to presentation. }
    \label{fig:vd_ury}
\end{figure}

\noindent Figure \ref{fig:vd_ury} shows the annual cycle of $v_d$ and $-\partial_y u_r$ as a zonal mean over the different regions. The annual cycle of zonally averaged $v_d$ is a near perfect sinusoid with an amplitude of $2 ms^{-1}$. The zonally averaged $-\partial_y u_r$ is simply the zonally averaged relative vorticity and it's annual cycle is orthogonal to that of $[v_d]$, though it's magnitude is diminished in the winter compared to summer. This arrangement of these two quantities hints at advection of vorticity from the summer hemisphere into the winter hemisphere by the divergent mean meridional wind \citep{hoskins2019detailed} and the resultant year-round southward flux of absolute vorticity \citep{zurita2019role}.
In fact, this theme of cross-equatorial transport of vorticity by the thermally direct circulation is consistent across all sectors and explains why there exists a strong correlation between the mean convergence computed over individual regions and the zonally averaged mean convergence (Table \ref{table:correlation}). The larger magnitude of zonal mean vorticity in the summer leads to the single seasonal peak in the mean flux convergence.

\noindent In Figure \ref{fig:rotDiv}, the zonal mean eddy momentum flux convergence is dominated by the $u_r v_d$ term because of the prominence of the thermally forced stationary Rossby waves in the tropics \citep{zurita2019role}. However, the picture is quite different in the localised sectors. In the AWP and CP-WA regions, the two eddy convergence terms (solid and dashed blue lines) align only around the spring season. This implies that the other components of the total eddy momentum flux die down during the spring and become important during other times of the year. Amongst all the sectors, the largest  contribution from the $u_r v_d$ eddy convergence is in the AWP sector. The CP-WA sector is, in fact, dominated by the $u_r v_r$ component (not shown), which implies a role of extratropical eddies \citep{zurita2019role}. However, these two convergence terms line up quite nicely during the NH monsoon season in the African sector. This fact suggests that the prominent role of the $u_r v_d$ flux in the zonal mean eddy convergence results from internal compensations between other components of the eddy momentum flux.

\section{Spatial Eddy Momentum Fluxes and Regional Hadley Cells}

As we have seen, the nature of momentum fluxes varies with season and region. Given that these are likely to influence different longitudinal sectors in the tropics, here, we explore the spatial character of the eddy fluxes and the associated overturning cells in the tropics during the winter, summer and spring equinox seasons.

\subsection{Winter}

\begin{figure}[ht]
    \centering
    
    \includegraphics[width=0.8\linewidth]{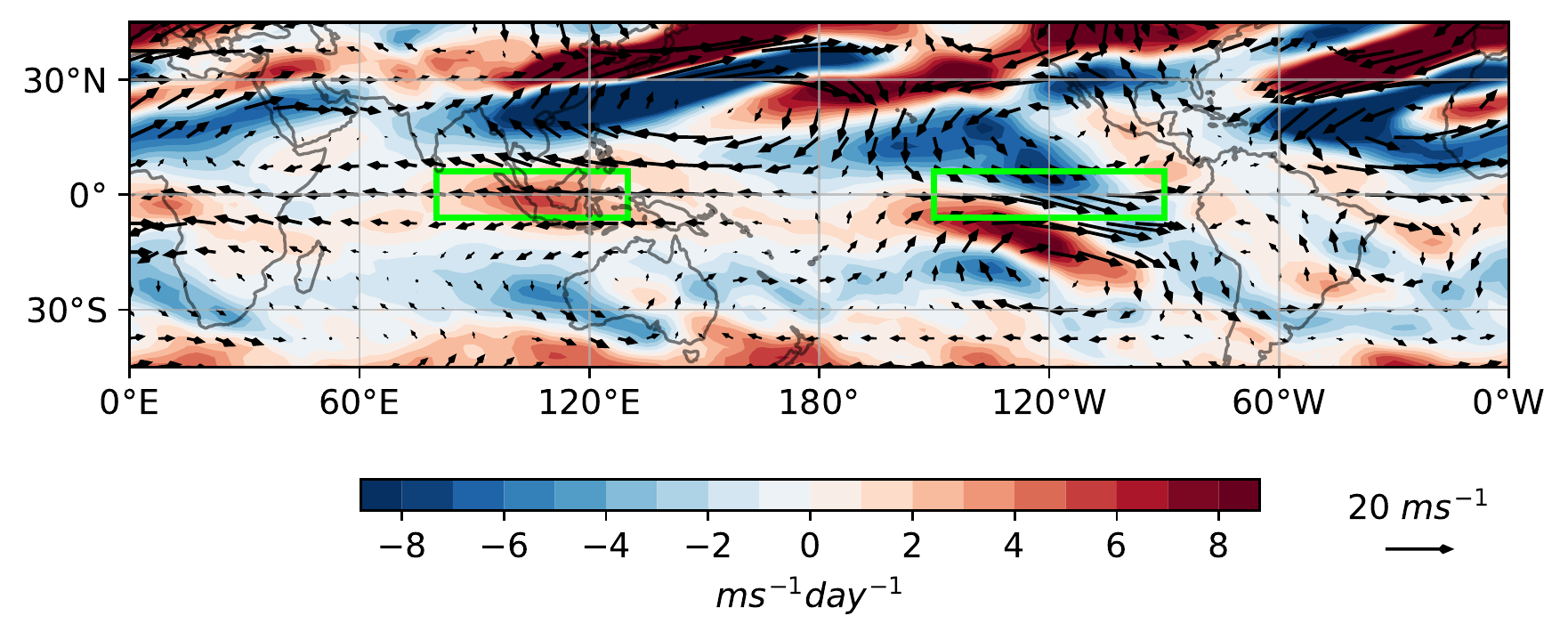}
    \caption{Spatial map of the upper tropospheric (150-300 hPa) eddy momentum flux convergence overlain with quivers of eddy wind vectors, seasonally averaged over winter. The quiver-key is shown in the bottom right of the figure. The green boxes are centred over the equator at 105E and 120W respectively and are $10^{\circ}$ wide in both cases. }
    \label{fig:eddiesClimWin}
\end{figure}

\noindent Figure \ref{fig:eddiesClimWin} shows the upper tropospheric spatial distribution of the eddy momentum flux convergence term seasonally averaged for winter. The aforementioned winter-time mismatch between the global and local domain estimates (solid blue lines in Figures \ref{fig:eddyVsMean} and \ref{fig:eddyVsMean3zones}) is evident here. When the eddy convergence is calculated globally, there is a cancellation between the positive contribution around 120E and the negative contribution from approximately 120W.
The features in Figure \ref{fig:eddiesClimWin} responsible for this mutual cancellation are highlighted by the green boxes. In fact, the eddy momentum flux divergence due to the 120W box can be seen in the dotted curve (CP-WA region) of Figure \ref{fig:eddyVsMean3zones}.
The positive eddy convergence in the 120E box is linked to the two off-equatorial anti-cyclonic Rossby gyres straddling the equator around the Maritime Continent --- see also the top panel of Figure 9 in \cite{dima2005tropical} --- of which the Southern Hemispheric (SH) gyre is tied to the Australian Monsoon. The eddy momentum convergence features associated with the 120W box are related to the Pacific Ocean upper tropospheric troughs \citep[][and the references therein]{kelly2016february}. These eddies can also be seen in the stationary wave patterns during the winter season \citep{held2002northern}. 

\noindent These regions of eddy momentum convergence/divergence can be explained using the eddy wind vectors ($u^*$,$v^*$). Considering Figure \ref{fig:eddiesClimWin}, in AWP the  combination of negative $u^*$ with negative (positive) $v^*$ southwards (northwards) of the Equator with zero $v^*$ at the Equator results in a northward (southward) eddy momentum flux to the south (north) of the Equator with zero flux at the Equator. This arrangement of eddy momentum fluxes leads to an accumulation of momentum at the Equator. For the region in the East Pacific, positive $u^*$ along with negative $v^*$ over most of the area gives a southward momentum flux. Moving away from the equator, this southward flux increases as $y$ (distance from equator, positive northwards) decreases and results in positive or almost zero convergence close to southern edge of the box.

\noindent The NH subtropical and extratropical features of Figure \ref{fig:eddiesClimWin} compare nicely with wave activity convergence features \citep{caballero2009impact}. This is testimony to the fact that a zone of eddy momentum flux convergence (divergence) is a zone of wave activity divergence (convergence). This is synonymous with the theory that Rossby wave breaking (convergence of wave activity) causes mean-flow deceleration. From their results, it is worthwhile to note that the divergence of eddy momentum flux in the 120W box of Figure \ref{fig:eddiesClimWin} is due to a weak convergence of wave activity emanating from SH (see also Figure S1). These waves, seemingly generated in SH Pacific ocean, propagate across the equator into NH through the "westerly window" in the 120W box \citep{hoskins1993rossby, li2015interhemispheric}. These waves travel northwards to join the waves coming in from Asia to form a region of confluence off the North American coast in the subtropics of eastern Pacific (Figure S1).Despite there being a pocket of westerlies in the equatorial Atlantic, the Rossby waves from the North Atlantic are absorbed in North Africa before they can reach the equator and there is no cross-equatorial propagation here \citep{caballero2009impact}.The cross-equatorial propagation observed during winter is not present in the other seasons (Figure S1).

\begin{figure}[ht]
    \centering
    
    \includegraphics[width=0.8\linewidth]{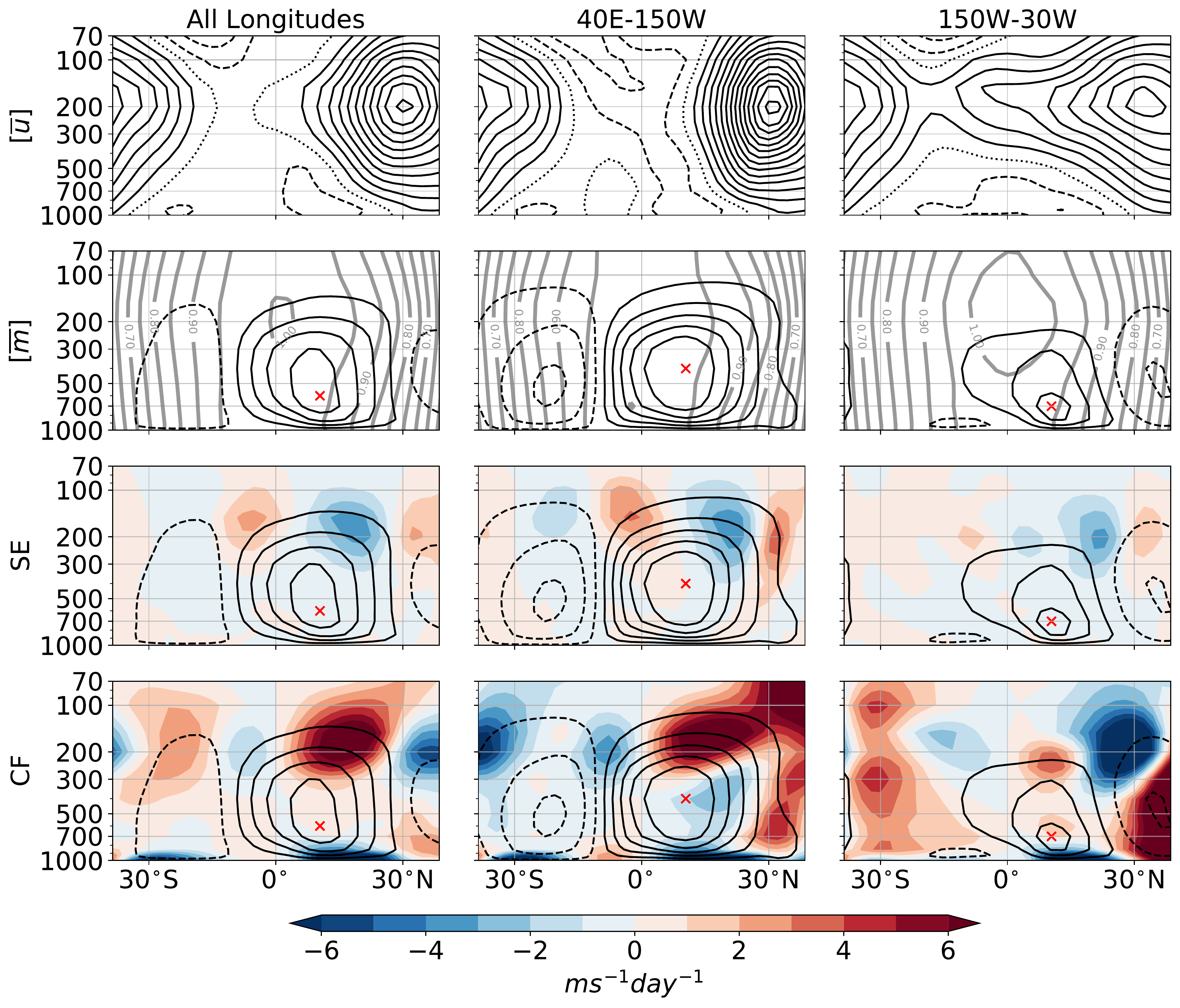}
    \caption{Winter time distribution of (top row) zonal wind, (second row) angular momentum per unit mass, (third row) stationary eddies and (fourth row) coriolis force computed over the global and local domains (Figure \protect \ref{fig:diffVd}) along with meridional overturning stream function (black contours, maximum stream function is marked for each domain by red cross). All quantities are presented as a zonal mean over the respective domains. In all panels, solid (dashed) contours indicate positive (negative) values and the zero contour is dotted (not provided for the overturning stream function). For $[\overline{u}]$, the contour interval is 4 $ms^{-1}$. For $[\overline{m}]$, heavier grey contours are used which are multiples of ${\Omega a^2}$ ($\Omega$ is the angular velocity of Earth about it's own axis and $a$ is its radius). For the overturning stream functions, a two point moving average has been applied along latitudes prior to presentation; positive (negative) values indicate clockwise (counter-clockwise) circulations and the contour interval is $5\times10^{10}$ $kgs^{-1}$.}
    \label{fig:totLocHadWin}
\end{figure} 

\noindent The top row of Figure \ref{fig:totLocHadWin} shows the vertical structure of zonally averaged zonal wind as a seasonal mean for winter. Similar to the global domain, the vertical structure of the zonal mean zonal wind over the AWP sector has mid-latitude westerlies and easterlies over most of the tropical atmosphere, except the weak surface westerlies near the equator in SH. Both domains show a maxima in the mid-latitude westerly winds in NH. Unlike these two domains, the CP-WA sector shows westerlies over the entire the upper troposphere. As noted before, this is also the westerly duct associated with cross-equatorial wave propagation. These westerlies are linked to the two eddies flanking the 120W box in Figure \ref{fig:eddiesClimWin}. Coupled with the weaker westerlies over tropical Atlantic, this is manifested in the form of weak equatorial superrotation in the zonal mean picture \citep{zurita2019role}.

\noindent In the second row, Figure \ref{fig:totLocHadWin} shows the zonally averaged meridional overturning stream function (black contours) over the respective sectors. These are computed using Equation 6 of \citet{zhang2013interannual}. Consistent with Figure \ref{fig:diffVd}, the AWP sector has stronger overturning circulations than the CP-WA region. The broad similarities between the zonally averaged quantities of global and 40E - 150W regions is evident here too, though the latter has stronger cells with greater meridional extent in both hemispheres and extends farther polewards and into the upper troposphere. The red marker in these panels denotes the position of the stream function maximum. It is clear that conventional HC is simply a area-weighted average of these localised overturning motions.

\noindent This discussion of individual cells colluding together to give the conventional HC is appropriated more importance once the zonally averaged angular momentum (second row of Figure \ref{fig:totLocHadWin}) is considered. Angular momentum per unit mass is expressed as $m = (\Omega a \cos{\phi} + u) a \cos{\phi}$, where $\phi$ is latitude. From Figure \ref{fig:totLocHadWin}, the solitary upper tropospheric angular momentum maxima above the equator in the global domain comes from the CP-WA sector. It is noteworthy that this maximum in angular momentum is tied to the weak superrotation identified in the previous paragraph and that it is a result of the zonal gradients in geopotential that exist in this region in the annual mean (Figure 4 of \cite{dima2005tropical}).
It should also be noted that the flow streamlines cross $[\overline{m}]$ contours, immediately suggesting a strong influence of the eddy flux in this region \citep{walker2006eddy}. In the AWP region, the rising branch of the HC in the SH and the upper tropospheric cross-equatorial flow does not cross $[\overline{m}]$ contours suggesting the conservation of angular momentum and a lack of eddy influence in this region. In both regions, as well as over the entire globe, the deviation of $[\overline{m}]$ contours from the vertical at the poleward extremity of the HC in the NH indicates a regime with vorticity based Rossby number between 0 and 1 \citep{walker2006eddy}.

\noindent It is well known that the winter HC is influenced by eddies \citep{caballero2007role, zurita2018coupled}, but here we have shown that the most significant affect of the eddies is in the 150W-30W (CP-WA) region. 
In fact, the AWP sector has the strongest overturning circulation and a large eddy momentum flux convergence but is not a region of strong eddy influence. In particular, the AWP winter cell is shielded from the SH eddies due to the band of upper level tropical easterlies as seen in Figure \ref{fig:eddiesClimWin} \citep{walker2006eddy}. 
As seen, this allows its rising branch to be nearly angular momentum conserving and to extend farther into the winter hemisphere \citep{schneider2008eddy}. However, the NH midlatitude eddies do exert an influence on AWP winter cell. This is via a deceleration of the subtropical jet due to the breaking of these waves on the poleward flank of this cell \citep{becker2001interaction, held2002northern}.
 
\noindent The dominant terms in Equations \ref{eq:2} and \ref{eq:3}, during winter are the stationary eddy flux in Equation \ref{eq:2} and the Coriolis term in Equation \ref{eq:3}; these are presented in the third and fourth rows of Figure \ref{fig:totLocHadWin}. 
The tropical eddy momentum convergence in this season is largely due to the stationary wave contribution from AWP. As discussed previously, this is because of the presence of the stationary Rossby waves straddling the Maritime Continent in the West Pacific. In fact, this large eddy momentum flux convergence during winter in this sector is seen in Figure \ref{fig:eddyVsMean3zones}.
In comparison the stationary eddy features in the CP-WA are much weaker and maximize at a lower level (200 hPa) than the AWP features. The convergence of eddy momentum flux (divergence of wave activity) from the Pacific Ocean upper tropospheric troughs in SH and the associated tropical and subtropical divergences (convergences) in NH, discussed previously, are evident here.
Also, the eddy momentum convergences (divergences) that occur because of the generation (absorption) of extratropical Rossby waves in the midlatitudes are visible here northwards (southwards) of $30^{\circ}$N in both the longitudinally limited sectors \citep{held2002northern, caballero2009impact}. 
In contrast, the Coriolis force has a major role to play across all longitudinal regions. In fact, in the deep tropics, it is the Coriolis force alone that balances the positive acceleration due to the stationary eddies in the AWP sector. Though, in the NH subtropics, the balance is largely between Coriolis, MMC and SE terms \citep{shaw2014role}. 

\noindent From the discussion so far, it is clear that the contrasting nature of eddy momentum convergence in the AWP and CP-WA sectors, observed in Figure \ref{fig:eddyVsMean3zones}, can be attributed to their respective origins. Reiterating, the positive acceleration in AWP sector is due to the convergence of westerly momentum by the tropical Rossby waves there. The deceleration in the CP-WA sector is due to the absorption/breaking of the waves coming towards the equator from the midlatitudes. The rather weak response (\textasciitilde$-0.5$ $ms^{-1}{day}^{-1}$) of the mean momentum convergences during this season is due to the feeble and symmetric nature of the Coriolis force close to the equator and because the non-linear momentum advection by the MMC term takes effect somewhat northward of the equator during this season. The CP-WA region, with a weak meridional gradient of zonal winds, also has a small MMC contribution. 

\subsection{Summer}

\begin{figure}[ht]
    \centering
    
    \includegraphics[width=0.8\linewidth]{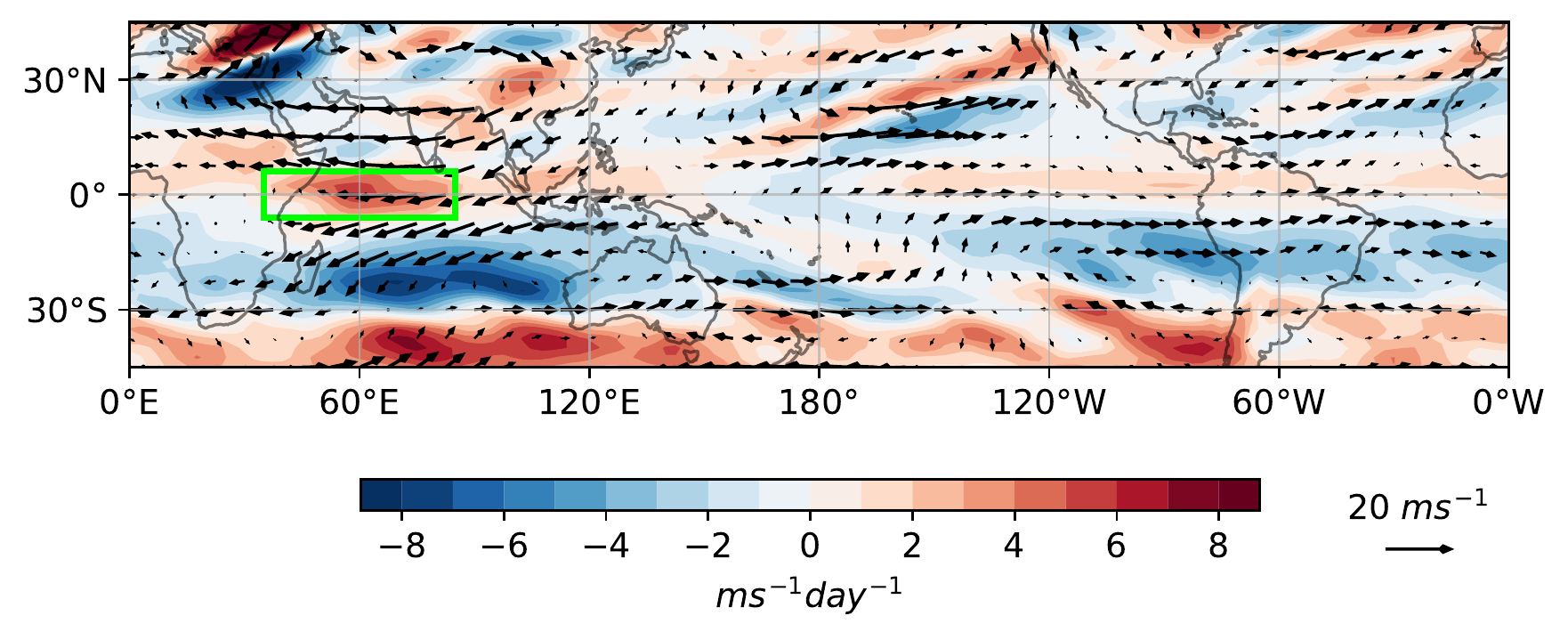}
    \caption{Same as Figure \protect \ref{fig:eddiesClimWin} except for summer. The green box is centered over the equator at 60E. }
    \label{fig:eddiesClimSum}
\end{figure}

\noindent Moving to the boreal summer, Figure \ref{fig:eddiesClimSum} shows the spatial distribution of the upper tropospheric eddy momentum flux convergence during this season. The dominant feature, highlighted by the box in Figure \ref{fig:eddiesClimSum}, responsible for the eddy flux convergence in the deep tropics during this season in Figure \ref{fig:eddyVsMean} (see also Figure \ref{fig:eddyVsMean3zones}) is the upper tropospheric return flow of the Asian summer monsoon \citep{dima2005tropical, hoskins2019detailed}. As discussed previously for winter, similar arguments can be made here regarding the eddy momentum flux coupled with the nature of the meridional divergence operator to explain the eddy flux convergence over the tropical Indian Ocean. In comparison to winter, the Pacific Ocean gyres have a much weaker tilt over the East Pacific sector during winter. This results in the weak contribution to the zonally averaged momentum budget from this region, consistent with the dotted blue lines in Figure \ref{fig:eddyVsMean3zones}.

\begin{figure}[ht]
    \centering
    
    \includegraphics[width=0.8\linewidth]{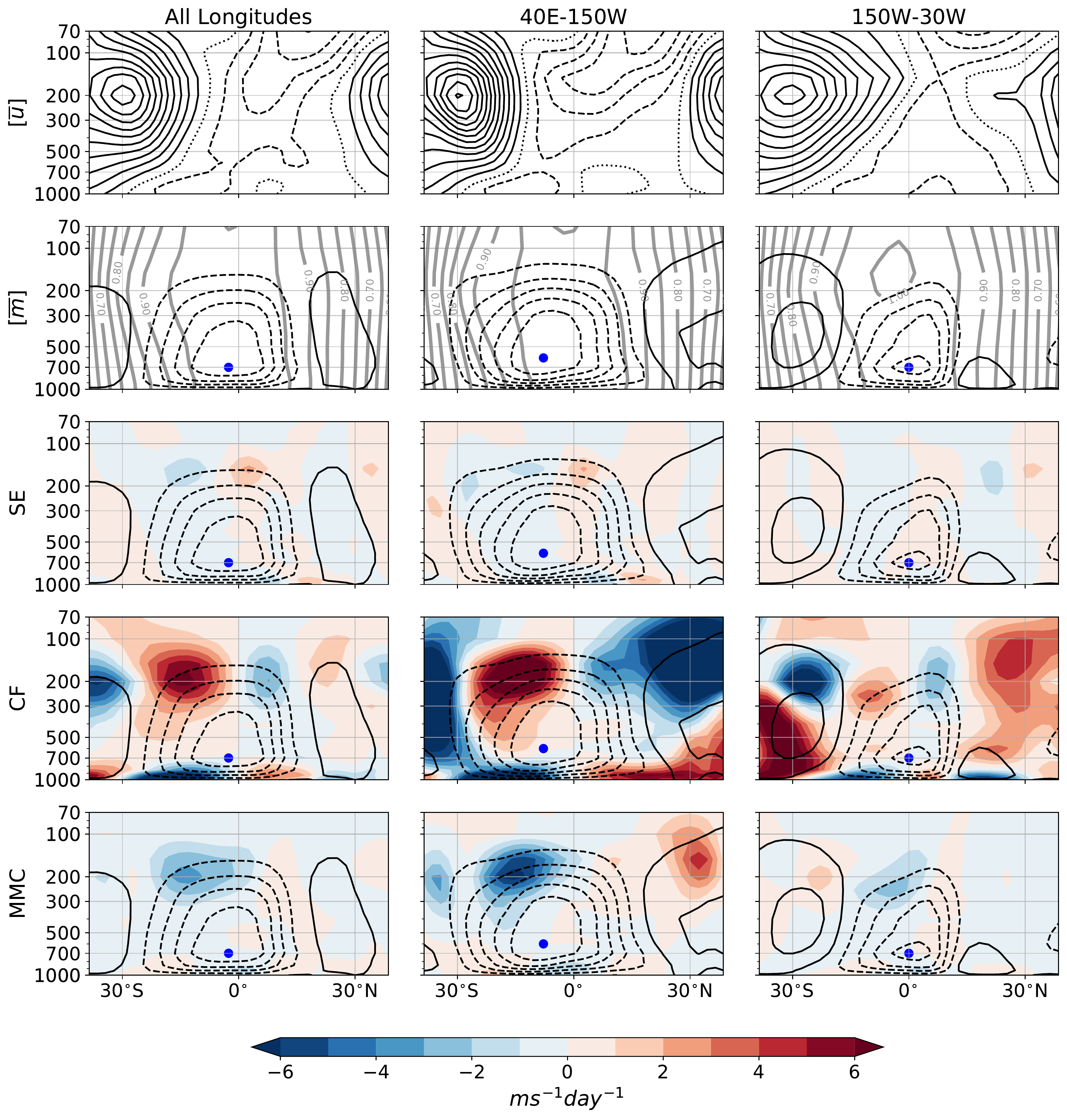}
    \caption{Same as Figure \protect \ref{fig:totLocHadWin} except for summer. The blue dots denote the mimima of the overturning stream function.
    }
    \label{fig:totLocHadSum}
\end{figure}

\noindent Figure \ref{fig:totLocHadSum} shows the vertical structure of the zonally averaged diagnostics for summer. The top row shows that the equatorial upper troposphere is dominated by easterlies which extend up to the subtropics in NH. As in Figure \ref{fig:totLocHadWin} for winter, the AWP sector has similar but stronger features in comparison to the global domain. Both these panels show the near-surface westerlies in tropical NH that are characteristic of the lower-level monsoonal flow. In contrast to winter, the CP-WA sector also shows equatorial upper tropospheric easterlies during summer. However, there does appear to be a hint of weak equatorial westerly winds over AWP (CP-WA) sector at 70 (150) hPa level.

\noindent The regional stream function and angular momentum contours during summer, shown in the second row of Figure \ref{fig:totLocHadSum}, are quite similar to their winter season counterparts, though the globally averaged HC has a different character.
Specifically, in the deep tropics the upper branches of winter hemisphere cross-equatorial cells in the global and AWP domains do not cross angular momentum contours and conserve angular momentum \citep{hoskins2019detailed, bordoni2008monsoons}. 
Now, a band of tropical easterlies shield the rising branch of the local cell from the influence of NH midlatitude eddies propagating towards the tropics \citep{bordoni2008monsoons, schneider2008eddy}.
Once again, there is an isolated angular momentum maximum in the CP-WA sector, and another one in the stratosphere of the AWP sector. 
In both the longitudinally limited sectors, these isolated maxima appear to be tied to the weak signature of the equatorial westerly winds. 
The cross-equatorial CP-WA cell does cross the isolated $[\overline{m}]$ contour at 200 hPa in the SH deep tropics indicating an eddy influence in this sector; however, this affect doesn't appear to carry through to the globally averaged HC. 
Thus, the upper branch of globally averaged cross-equatorial HC does not cross angular momentum contours and is free of eddy influence \citep{hoskins2019detailed, walker2006eddy}, which is in contrast to the global cell in the winter season.

\noindent As discussed with regard to Figure \ref{fig:eddiesClimSum} and evident in the third row of Figure \ref{fig:totLocHadSum}, the dominant contribution to the eddy momentum flux convergence in the equatorial upper troposphere is by the monsoonal return flow of the AWP sector. The amplitudes of SE during this season are much reduced than that observed in winter. Though diminished with respect to the winter season, the stationary waves generated in the NH are still visible and the extratropical stationary eddy signature is dominated by the CP-WA sector. The NH subtropical region in AWP that was observed to be a zone of strong eddy momentum flux divergence during winter, is one of weak convergence during summer. The lower strength of these waves is probably accounted for by the weakened NH subtropical jet and that the NH meridional temperature gradient is reduced during this season \citep{held2002northern, lee2003dynamical}. The transient eddy contribution is dominantly via waves from the midlatitudes of the SH \citep{ambrizzi1995rossby}. The features of eddy momentum convergence/divergence observed in Figure \ref{fig:eddiesClimSum} around $30^\circ$S are signatures of these transient eddies. They are small in the tropics with respect to SE and hence, left out of consideration here. 

\noindent Similar to the winter calculations, here too the Coriolis force plays a major role. Consistent with the observations of \citet{shaw2014role}, the dominant balance in the deep tropics is now between three terms; CF, MMC and SE.
Towards the SH subtropics, the balance is between Coriolis, MMC and TE (not shown) with a relatively small contribution from SE. 
\citet{bordoni2008monsoons} note the importance of high frequency transient eddies in this balance with a negligible role for the stationary eddies, before and after the monsoon onset over the monsoon sector.
The near-surface balance in the tropics is between Coriolis and friction \citep{shaw2014role}, dominated by the AWP sector. This explains the NH tropical near-surface westerlies. As during winter, the MMC has one dominant feature in the subtropics of the winter hemisphere and it is controlled by the AWP sector. Though, near the equator, the MMC flux convergence in the CP-WA region is actually greater in magnitude than that in the AWP region (Figure \ref{fig:eddyVsMean3zones}).

\noindent Putting the tendencies of summer season (Figure \ref{fig:eddyVsMean3zones}) in perspective, the mean meridional momentum flux divergence from AWP is much smaller than the other two sectors. As the AWP sector has weaker $-\partial_y[\overline{u}]$ in the deep tropics despite having stronger overturning circulation ($[\overline{v}]$), the MMC contribution from this region gains traction farther away from the equator. This coupled with the non-existent Coriolis term close to the equator explains the weak mean meridional flux divergence linked to this sector during summer. However, the CP-WA sector has a larger zonal wind gradient close to the equator which accounts for its stronger MMC response. With regard to the eddy fluxes, they are almost entirely accounted for by the AWP region and their tropical origin leads to an accelerating tendency on the zonal mean flow.

\subsection{Spring}

\begin{figure}[ht]
    \centering
    \includegraphics[width=0.8\linewidth]{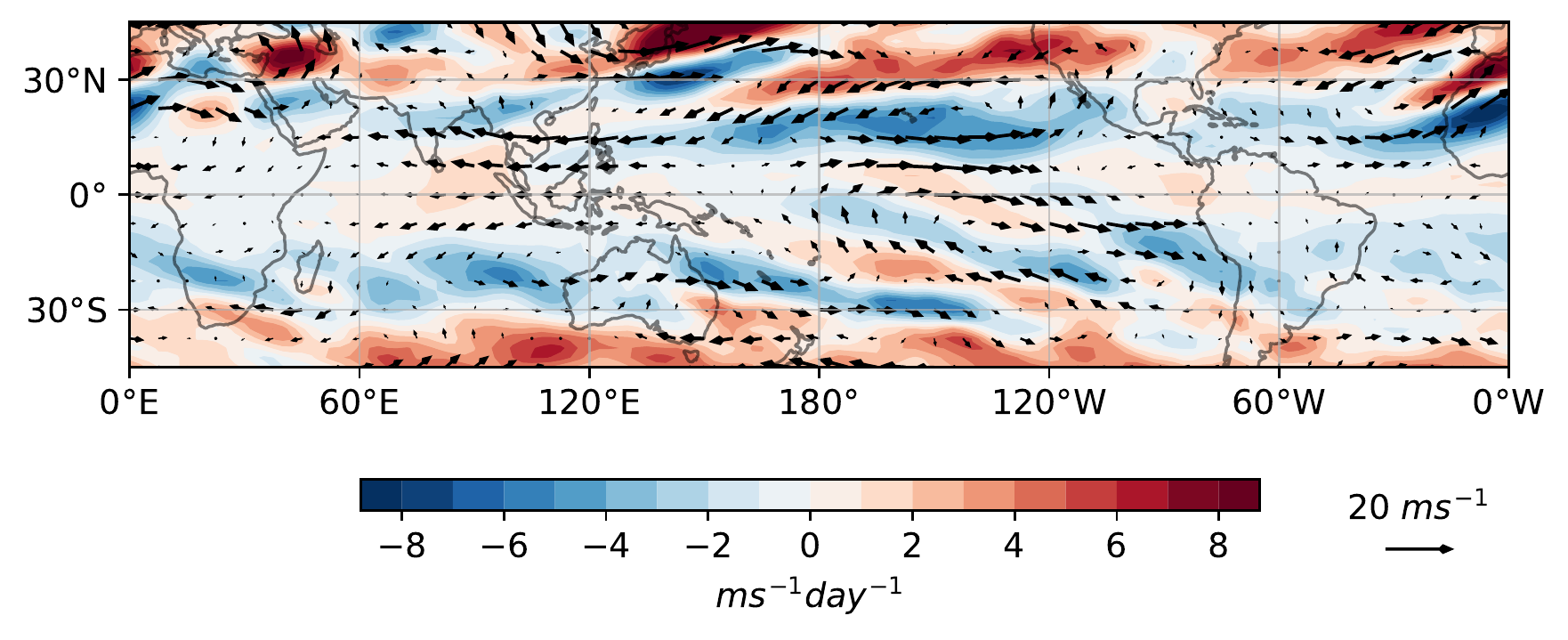}
    \caption{Same as Figure \protect \ref{fig:eddiesClimWin} except for spring.}
    \label{fig:eddiesClimSpring}
\end{figure}

\noindent The features of the equinoctial spring season can be seen as a transition from NH winter to NH summer. Figure \ref{fig:eddiesClimSpring} shows the spatial features of the equatorial upper troposphere as a seasonal average for spring. On comparison with Figure \ref{fig:eddiesClimWin} and Figure \ref{fig:eddiesClimSum}, it is apparent that there is a weakening of the eddy momentum flux convergence over the Maritime Continent and gradual strengthening over the tropical Indian Ocean. The strong eddy momentum flux convergence/ divergence features observed near 120W in the Pacific (Figure \ref{fig:eddiesClimWin}) has decreased to an intermediate value. This is due to the decreasing tilt of the Pacific Ocean subtropical gyres as the seasons progress from winter to summer, reaffirmed by the near zero eddy momentum convergence in CP-WA (blue dotted lines in Figure \ref{fig:eddyVsMean3zones}). Apart from these features in the deep tropics, there are considerable differences in the subtropics. The decreasing strength of the NH stationary eddies in the Pacific and Atlantic Ocean and increasing strength of the SH baroclinic zones is clear. 

\begin{figure}[ht]
    \centering
    \includegraphics[width=0.8\linewidth]{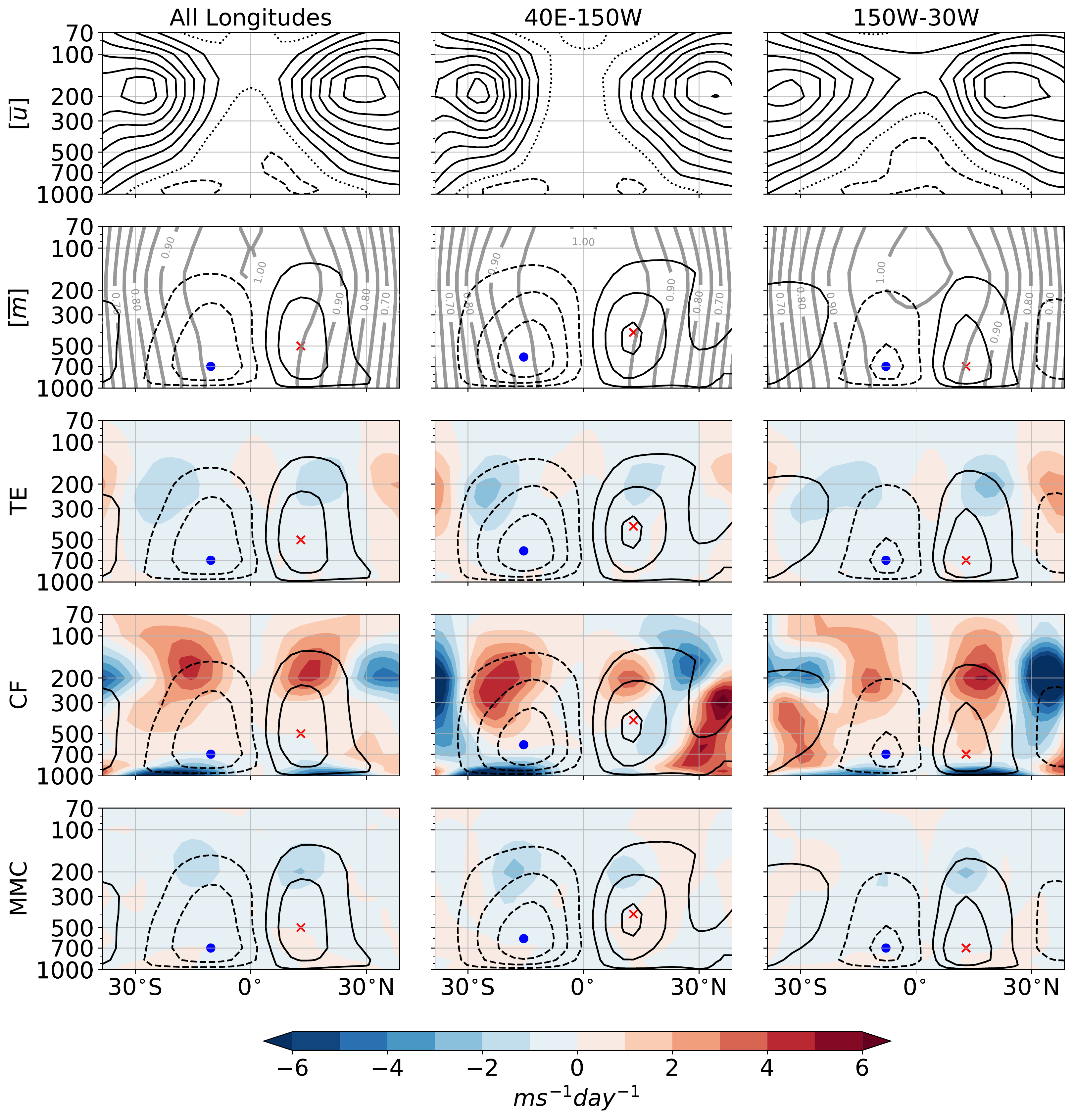}
    \caption{Same as Figure \protect \ref{fig:totLocHadWin} except for spring. The blue dots (red crosses) denote the mimima (maxima) of the meridional overturning stream function.
    }
    \label{fig:totLocHadSpring}
\end{figure}

\noindent Figure \ref{fig:totLocHadSpring} shows the vertical structure of the zonal mean diagnostics seasonally averaged over spring. In comparison to the winter and summer, the zonal mean zonal winds are more symmetric about the equator. As in the winter season, the CP-WA sector shows upper tropospheric westerly winds  contributed by the Pacific ocean upper tropospheric troughs, though they are much weaker. The AWP sector has easterlies throughout the depth of the equatorial troposphere. The globally averaged zonal winds exhibit a weak superrotation limited to 150 hPa level and above.

\noindent The meridional overturning cells show equatorial symmetry across all regions, although the region of ascent is displaced slightly into NH \citep{dima2005tropical}.
The NH cells extend slightly deeper into the troposphere.
There is an isolated maximum of angular momentum away from the surface in the global domain as well as the CP-WA estimate. 
The deviation of the angular momentum contours from the vertical in all sectors implies a regime intermediate between angular momentum conserving and non-conserving schemes \citep{walker2006eddy}. 
This implies a greater role for the eddies across all sectors in comparison to the solstitial seasons.

\noindent From the discussions related to Figure \ref{fig:eddyVsMean}, we know that both the mean meridional and eddy momentum flux convergences in the deep tropics are near-zero during the equinoctial seasons. This is evident from the subsequent panels in Figure \ref{fig:totLocHadSpring}. The dominant balance during this season is between the subtropical features of TE, CF and MMC. Compared to the two solstitial seasons, the SE term has a inconsequential role here and it's contribution in the deep tropics is similar to TE. 
It should also be noted that the mild westerly acceleration by the TE is also observed during the other seasons but it gets overshadowed by the strong SE contribution of the climatological Rossby gyres.

\noindent In all, the spring season showcases a high degree of equatorial symmetry marked by a pair of nearly symmetric overturning cells (Figure \ref{fig:totLocHadSpring}).
This is due to a lack of eddy forcing of equatorial Rossby waves and the primary thermal forcing during this season is the incoming solar radiation \citep{norton2006tropical}.
This is evident here due the lack of Rossby gyres (Figure \ref{fig:eddiesClimSpring}) seen during the two solstitial seasons (Figure \ref{fig:eddiesClimWin} and \ref{fig:eddiesClimSum}). 
This lack of eddy forcing is evident in Figure \ref{fig:eddyVsMean} with eddy momentum flux convergence $\leq 0.2$ $ms^{-1}{day}^{-1}$ during this season from Day 100-150.
Interestingly, the mean meridional flux convergence provides a weak acceleration during this period. 
This happens because of the non-zero Coriolis effect on the southern cell as it's rising edge is positioned over the equator (Figure \ref{fig:eddiesClimSpring}).

\section{Conclusions and Discussion}

As discussed in the Introduction, the zonal mean tropical eddy and mean momentum fluxes and the behaviour of the conventional Hadley cell is fairly well documented in literature. 
Here, we have focused on the temporal structure of these zonal mean entities and also studied the longitudinal structure underlying them. At the outset, the zonal mean eddy and mean fluxes are shown to have a pronounced annual cycle that peaks in the boreal summer. Further, the deep tropical fluxes almost vanish during times of the equinox. The nature of these fluxes is probed further by splitting the globe into three regions based on the divergent motions forced by the longitudinally heterogeneous monsoon heating. Specifically, these are the Asia-West Pacific (AWP), central Pacific-West Atlantic (CP-WA) and African sectors. 

\noindent The AWP sector provides the bulk of the eddy momentum flux convergence during both the solstitial seasons via thermally forced stationary eddies in the tropics, while the CP-WA sector receives momentum flux during winter from extratropical eddies propagating into the tropics from the southern hemisphere via the East Pacific. The disparate origin of the eddies in these two regions causes the AWP and CP-WA regions to experience acceleration and deceleration of the zonal flow, respectively. The African sector shows relatively low levels of eddy flux through the year. Thus, the nature of the eddy flux in individual sectors is very different from the zonal mean perspective. In fact,  the prominent contribution of rotational zonal - divergent meridional component to the zonal mean eddy convergence \citep{zurita2019role}, is also not observed to hold in these isolated sectors.

\noindent Although the AWP sector dominates the eddy convergence in the deep tropics, it comes up short with respect to the contribution towards the mean convergence. This is due to the fact that although the meridional divergent winds are comparable across all sectors, the meridional gradient of the zonal wind is lower in the deep tropics for AWP; however it picks up farther into the winter hemisphere as the AWP local overturning cell outflow merges with the subtropical jet. In fact, it is the African and CP-WA regions that contribute most significantly to the mean convergence in the deep tropics. But, the qualitative nature of the mean convergence is remarkably cohesive across all three regions and this is traced to the  advection of absolute vorticity by the divergent meridional wind in localized cross-equatorial cells.
Indeed, the flux has a single peak in the boreal summer due to the larger zonal mean vorticity in this season.
Further, the high degree of compensation observed between the mean meridional and eddy momentum flux convergence in the zonal mean \citep{dima2005tropical,kraucunas2007tropical} does not hold in the individual sectors. 

\noindent The zonal mean overturning circulation is dominated by AWP cell \citep{hoskins2019detailed}.
It is seen that the AWP cell is thermally-direct during both the solstitial seasons while the CP-WA cell is eddy-driven to varying degrees. In fact, the CP-WA region shows an isolated angular momentum extremum in the upper tropopshere through the year indicating significant eddy influence.
This is also reflected in the zonal mean HC. For the most part, extratropical eddies are able to reach the deep tropics at upper levels only during winter and remain confined to the outer reaches of the HC during summer. In all, the summer and winter balance of fluxes is between the Coriolis, stationary eddy and to a lesser extent the mean meridional contribution. Whereas, in the spring equinox, the deep tropics show very low levels of momentum flux convergence. Thus, the zonal mean HC is an aggregation of the localised responses to thermal and eddy driving of the overturning circulation.

\noindent These results show the extent of internal cancellations which occur when momentum flux convergence is integrated over all longitudes. Indeed, the compensation between the eddy and mean meridional flux does not actually hold in longitudinal zones. Moreover, a Helmholtz decomposition suggests that the mean flux is indeed controlled by the rotational zonal and divergent meridional component, but this is not true for the eddy flux when examined in individual sectors. Further, the influence of midlatitudes on the overturning tropical circulation is seen to be most prominent in the CP-WA region. The spatial structure put forth in this work not only helps in uncovering the heterogeneous nature of the deep tropical momentum flux budget and regions of influence of extratropical eddies, it also shows the delicate longitudinal cancellations in the present day climate that give rise to the familiar zonal mean picture of the tropics.

\section*{Acknowledgements}
The authors would like to thank the Divecha Centre at IISc for financial assistance. 

\bibliographystyle{apalike}
\bibliography{References.bib}

\begin{thebibliography}{}

\bibitem[Adler et~al., 2003]{adler2003version}
Adler, R.~F., Huffman, G.~J., Chang, A., Ferraro, R., Xie, P.-P., Janowiak, J.,
  Rudolf, B., Schneider, U., Curtis, S., Bolvin, D., Gruber, A., Susskind, J.,
  Arkin, P., and Nelkin, E. (2003).
\newblock {The Version-2 Global Precipitation Climatology Project (GPCP)
  Monthly Precipitation Analysis (1979–Present)}.
\newblock {\em Journal of Hydrometeorology}, 4(6):1147--1167.

\bibitem[Ambrizzi et~al., 1995]{ambrizzi1995rossby}
Ambrizzi, T., Hoskins, B.~J., and Hsu, H.-H. (1995).
\newblock {Rossby Wave Propagation and Teleconnection Patterns in the Austral
  Winter}.
\newblock {\em Journal of the Atmospheric Sciences}, 52(21):3661--3672.

\bibitem[Becker and Schmitz, 2001]{becker2001interaction}
Becker, E. and Schmitz, G. (2001).
\newblock {Interaction between Extratropical Stationary Waves and the Zonal
  Mean Circulation}.
\newblock {\em Journal of the Atmospheric Sciences}, 58(5):462--480.

\bibitem[Becker et~al., 1997]{becker1997}
Becker, E., Schmitz, G., and Geprags, R. (1997).
\newblock The feedback of midlatitude waves onto the hadley cell in a simple
  general circulation model.
\newblock {\em Tellus A}, 49(2):182--199.

\bibitem[Boos and Emanuel, 2008]{booseman}
Boos, W.~R. and Emanuel, K.~A. (2008).
\newblock Wind{\textendash}evaporation feedback and abrupt seasonal transitions
  of weak, axisymmetric hadley circulations.
\newblock {\em Journal of the Atmospheric Sciences}, 65(7):2194--2214.

\bibitem[Bordoni and Schneider, 2008]{bordoni2008monsoons}
Bordoni, S. and Schneider, T. (2008).
\newblock Monsoons as eddy-mediated regime transitions of the tropical
  overturning circulation.
\newblock {\em Nature Geoscience}, 1(8):515--519.

\bibitem[Bordoni and Schneider, 2010]{bordoni-schneider2010}
Bordoni, S. and Schneider, T. (2010).
\newblock Regime transitions of steady and time-dependent hadley circulations:
  Comparison of axisymmetric and eddy-permitting simulations.
\newblock {\em Journal of the Atmospheric Sciences}, 67(5):1643--1654.

\bibitem[Caballero, 2007]{caballero2007role}
Caballero, R. (2007).
\newblock Role of eddies in the interannual variability of hadley cell
  strength.
\newblock {\em Geophysical Research Letters}, 34(22).

\bibitem[Caballero and Anderson, 2009]{caballero2009impact}
Caballero, R. and Anderson, B.~T. (2009).
\newblock Impact of midlatitude stationary waves on regional hadley cells and
  {ENSO}.
\newblock {\em Geophysical Research Letters}, 36(17).

\bibitem[Davis and Birner, 2019]{Davis}
Davis, N.~A. and Birner, T. (2019).
\newblock Eddy influences on the hadley circulation.
\newblock {\em Journal of Advances in Modeling Earth Systems},
  11(6):1563--1581.

\bibitem[Dee et~al., 2011]{dee2011era}
Dee, D.~P. et~al. (2011).
\newblock The era-interim reanalysis: configuration and performance of the data
  assimilation system.
\newblock {\em Quarterly Journal of the Royal Meteorological Society},
  137(656):553--597.

\bibitem[Dima et~al., 2005]{dima2005tropical}
Dima, I.~M., Wallace, J.~M., and Kraucunas, I. (2005).
\newblock Tropical zonal momentum balance in the {NCEP} reanalyses.
\newblock {\em Journal of the Atmospheric Sciences}, 62(7):2499--2513.

\bibitem[Emanuel, 1995]{eman95}
Emanuel, K.~A. (1995).
\newblock {On Thermally Direct Circulations in Moist Atmospheres}.
\newblock {\em Journal of the Atmospheric Sciences}, 52(9):1529--1534.

\bibitem[Fang and Tung, 1999]{fangtung}
Fang, M. and Tung, K.~K. (1999).
\newblock {Time-Dependent Nonlinear Hadley Circulation}.
\newblock {\em Journal of the Atmospheric Sciences}, 56(12):1797--1807.

\bibitem[Frierson et~al., 2007]{frierson}
Frierson, D. M.~W., Lu, J., and Chen, G. (2007).
\newblock Width of the hadley cell in simple and comprehensive general
  circulation models.
\newblock {\em Geophysical Research Letters}, 34(18).

\bibitem[Geen et~al., 2018]{geen}
Geen, R., Lambert, F.~H., and Vallis, G.~K. (2018).
\newblock Regime change behavior during asian monsoon onset.
\newblock {\em Journal of Climate}, 31(8):3327--3348.

\bibitem[Gill, 1980]{gill1980}
Gill, A.~E. (1980).
\newblock Some simple solutions for heat-induced tropical circulation.
\newblock {\em Quarterly Journal of the Royal Meteorological Society},
  106(449):447--462.

\bibitem[Held and Hou, 1980]{held1980nonlinear}
Held, I.~M. and Hou, A.~Y. (1980).
\newblock {Nonlinear Axially Symmetric Circulations in a Nearly Inviscid
  Atmosphere}.
\newblock {\em Journal of the Atmospheric Sciences}, 37(3):515--533.

\bibitem[Held et~al., 2002]{held2002northern}
Held, I.~M., Ting, M., and Wang, H. (2002).
\newblock {Northern Winter Stationary Waves: Theory and Modeling}.
\newblock {\em Journal of Climate}, 15(16):2125--2144.

\bibitem[Hoskins and Ambrizzi, 1993]{hoskins1993rossby}
Hoskins, B.~J. and Ambrizzi, T. (1993).
\newblock {Rossby Wave Propagation on a Realistic Longitudinally Varying Flow}.
\newblock {\em Journal of the Atmospheric Sciences}, 50(12):1661--1671.

\bibitem[Hoskins et~al., 2020]{hoskins2019detailed}
Hoskins, B.~J., Yang, G.-Y., and Fonseca, R.~M. (2020).
\newblock The detailed dynamics of the june{\textendash}august hadley cell.
\newblock {\em Quarterly Journal of the Royal Meteorological Society},
  146(727):557--575.

\bibitem[Kelly and Mapes, 2011]{kelly2011zonal}
Kelly, P. and Mapes, B. (2011).
\newblock Zonal mean wind, the indian monsoon, and july drying in the western
  atlantic subtropics.
\newblock {\em Journal of Geophysical Research: Atmospheres}, 116(D21).

\bibitem[Kelly and Mapes, 2016]{kelly2016february}
Kelly, P. and Mapes, B. (2016).
\newblock February drying in southeastern brazil and the australian monsoon:
  Global mechanism for a regional rainfall feature.
\newblock {\em Journal of Climate}, 29(20):7529--7546.

\bibitem[Kim and Lee, 2001]{kimlee}
Kim, H.-k. and Lee, S. (2001).
\newblock {Hadley Cell Dynamics in a Primitive Equation Model. Part II:
  Nonaxisymmetric Flow}.
\newblock {\em Journal of the Atmospheric Sciences}, 58(19):2859--2871.

\bibitem[Korty and Schneider, 2008]{korty}
Korty, R.~L. and Schneider, T. (2008).
\newblock Extent of hadley circulations in dry atmospheres.
\newblock {\em Geophysical Research Letters}, 35(23).

\bibitem[Kraucunas and Hartmann, 2005]{kraucunas2005equatorial}
Kraucunas, I. and Hartmann, D.~L. (2005).
\newblock Equatorial superrotation and the factors controlling the zonal-mean
  zonal winds in the tropical upper troposphere.
\newblock {\em Journal of the Atmospheric Sciences}, 62(2):371--389.

\bibitem[Kraucunas and Hartmann, 2007]{kraucunas2007tropical}
Kraucunas, I. and Hartmann, D.~L. (2007).
\newblock Tropical stationary waves in a nonlinear shallow-water model with
  realistic basic states.
\newblock {\em Journal of the Atmospheric Sciences}, 64(7):2540--2557.

\bibitem[Kuo, 1956]{kuo}
Kuo, H.-L. (1956).
\newblock {FORCED AND FREE MERIDIONAL CIRCULATIONS IN THE ATMOSPHERE}.
\newblock {\em Journal of Meteorology}, 13(6):561--568.

\bibitem[Lee, 1999]{lee1999climatological}
Lee, S. (1999).
\newblock {Why Are the Climatological Zonal Winds Easterly in the
  EquatorialUpper Troposphere?}
\newblock {\em Journal of the Atmospheric Sciences}, 56(10):1353--1363.

\bibitem[Lee and Kim, 2003]{lee2003dynamical}
Lee, S. and Kim, H.-k. (2003).
\newblock {The Dynamical Relationship between Subtropical and Eddy-Driven
  Jets}.
\newblock {\em Journal of the Atmospheric Sciences}, 60(12):1490--1503.

\bibitem[Li et~al., 2015]{li2015interhemispheric}
Li, Y., Li, J., Jin, F.~F., and Zhao, S. (2015).
\newblock Interhemispheric propagation of stationary rossby waves in a
  horizontally nonuniform background flow.
\newblock {\em Journal of the Atmospheric Sciences}, 72(8):3233--3256.

\bibitem[Lindzen and Hou, 1988]{lindzen1988hadley}
Lindzen, R.~S. and Hou, A.~V. (1988).
\newblock {Hadley Circulations for Zonally Averaged Heating Centered off the
  Equator}.
\newblock {\em Journal of the Atmospheric Sciences}, 45(17):2416--2427.

\bibitem[Nguyen et~al., 2017]{nguyen2018variability}
Nguyen, H., Hendon, H.~H., Lim, E.~P., Boschat, G., Maloney, E., and Timbal, B.
  (2017).
\newblock Variability of the extent of the hadley circulation in the southern
  hemisphere: a regional perspective.
\newblock {\em Climate Dynamics}, 50(1-2):129--142.

\bibitem[Norton, 2006]{norton2006tropical}
Norton, W.~A. (2006).
\newblock Tropical wave driving of the annual cycle in tropical tropopause
  temperatures. part {II}: Model results.
\newblock {\em Journal of the Atmospheric Sciences}, 63(5):1420--1431.

\bibitem[Plumb and Hou, 1992]{plumbhou}
Plumb, R.~A. and Hou, A.~Y. (1992).
\newblock {The Response of a Zonally Symmetric Atmosphere to Subtropical
  Thermal Forcing: Threshold Behavior}.
\newblock {\em Journal of the Atmospheric Sciences}, 49(19):1790--1799.

\bibitem[Raiter et~al., 2020]{raiter2020tropical}
Raiter, D., Galanti, E., and Kaspi, Y. (2020).
\newblock The tropical atmospheric conveyor belt: A coupled eulerian-lagrangian
  analysis of the large-scale tropical circulation.
\newblock {\em Geophysical Research Letters}, 47(10).

\bibitem[Satoh, 1994]{satoh}
Satoh, M. (1994).
\newblock {Hadley Circulations in Radiative–Convective Equilibrium in an
  Axially Symmetric Atmosphere}.
\newblock {\em Journal of the Atmospheric Sciences}, 51(13):1947--1968.

\bibitem[Schneider, 1977]{schn1977}
Schneider, E.~K. (1977).
\newblock {Axially Symmetric Steady-State Models of the Basic State for
  Instability and Climate Studies. Part II. Nonlinear Calculations}.
\newblock {\em Journal of the Atmospheric Sciences}, 34(2):280--296.

\bibitem[Schneider, 1984]{schneider1984response}
Schneider, E.~K. (1984).
\newblock {Response of the Annual and Zonal Mean Winds and Temperatures to
  Variations in the Heat and Momentum Sources}.
\newblock {\em Journal of the Atmospheric Sciences}, 41(7):1093--1115.

\bibitem[Schneider and Lindzen, 1977]{schneider1977axially}
Schneider, E.~K. and Lindzen, R.~S. (1977).
\newblock {Axially Symmetric Steady-State Models of the Basic State for
  Instability and Climate Studies. Part I. Linearized Calculations}.
\newblock {\em Journal of the Atmospheric Sciences}, 34(2):263--279.

\bibitem[Schneider and Bordoni, 2008]{schneider2008eddy}
Schneider, T. and Bordoni, S. (2008).
\newblock Eddy-mediated regime transitions in the seasonal cycle of a hadley
  circulation and implications for monsoon dynamics.
\newblock {\em Journal of the Atmospheric Sciences}, 65(3):915--934.

\bibitem[Schwendike et~al., 2014]{schwendike2014local}
Schwendike, J., Govekar, P., Reeder, M.~J., Wardle, R., Berry, G.~J., and
  Jakob, C. (2014).
\newblock Local partitioning of the overturning circulation in the tropics and
  the connection to the hadley and walker circulations.
\newblock {\em Journal of Geophysical Research: Atmospheres},
  119(3):1322--1339.

\bibitem[Shaw, 2014]{shaw2014role}
Shaw, T.~A. (2014).
\newblock On the role of planetary-scale waves in the abrupt seasonal
  transition of the northern hemisphere general circulation.
\newblock {\em Journal of the Atmospheric Sciences}, 71(5):1724--1746.

\bibitem[Singh et~al., 2017]{Singh}
Singh, M.~S., Kuang, Z., and Tian, Y. (2017).
\newblock Eddy influences on the strength of the hadley circulation: Dynamic
  and thermodynamic perspectives.
\newblock {\em Journal of the Atmospheric Sciences}, 74(2):467--486.

\bibitem[Sun et~al., 2018]{sun2019regional}
Sun, Y., Li, L. Z.~X., Ramstein, G., Zhou, T., Tan, N., Kageyama, M., and Wang,
  S. (2018).
\newblock Regional meridional cells governing the interannual variability of
  the hadley circulation in boreal winter.
\newblock {\em Climate Dynamics}, 52(1-2):831--853.

\bibitem[Walker and Schneider, 2006]{walker2006eddy}
Walker, C.~C. and Schneider, T. (2006).
\newblock Eddy influences on hadley circulations: Simulations with an idealized
  {GCM}.
\newblock {\em Journal of the Atmospheric Sciences}, 63(12):3333--3350.

\bibitem[Wang and Ting, 1999]{wang1999seasonal}
Wang, H. and Ting, M. (1999).
\newblock {Seasonal Cycle of the Climatological Stationary Waves in the
  NCEP–NCAR Reanalysis}.
\newblock {\em Journal of the Atmospheric Sciences}, 56(22):3892--3919.

\bibitem[Watterson and Schneider, 1987]{watterson1987effect}
Watterson, I.~G. and Schneider, E.~K. (1987).
\newblock The effect of the hadley circulation on the meridional propagation of
  stationary waves.
\newblock {\em Quarterly Journal of the Royal Meteorological Society},
  113(477):779--813.

\bibitem[Zhai and Boos, 2015]{ZhaiBoos}
Zhai, J. and Boos, W. (2015).
\newblock Regime transitions of cross-equatorial hadley circulations with
  zonally asymmetric thermal forcings.
\newblock {\em Journal of the Atmospheric Sciences}, 72(10):3800--3818.

\bibitem[Zhang and Wang, 2013]{zhang2013interannual}
Zhang, G. and Wang, Z. (2013).
\newblock Interannual variability of the atlantic hadley circulation in boreal
  summer and its impacts on tropical cyclone activity.
\newblock {\em Journal of Climate}, 26(21):8529--8544.

\bibitem[Zurita-Gotor, 2019]{zurita2019role}
Zurita-Gotor, P. (2019).
\newblock The role of the divergent circulation for large-scale eddy momentum
  transport in the tropics. part i: Observations.
\newblock {\em Journal of the Atmospheric Sciences}, 76(4):1125--1144.

\bibitem[Zurita-Gotor and {\'{A}}lvarez-Zapatero, 2018]{zurita2018coupled}
Zurita-Gotor, P. and {\'{A}}lvarez-Zapatero, P. (2018).
\newblock Coupled interannual variability of the hadley and ferrel cells.
\newblock {\em Journal of Climate}, 31(12):4757--4773.

\end{thebibliography}

% \section{Supplementary Information}
% \input{supporting.tex}

\end{document}

\maketitle

\begin{figure}[h]
    \centering
    \includegraphics[width=0.8\linewidth]{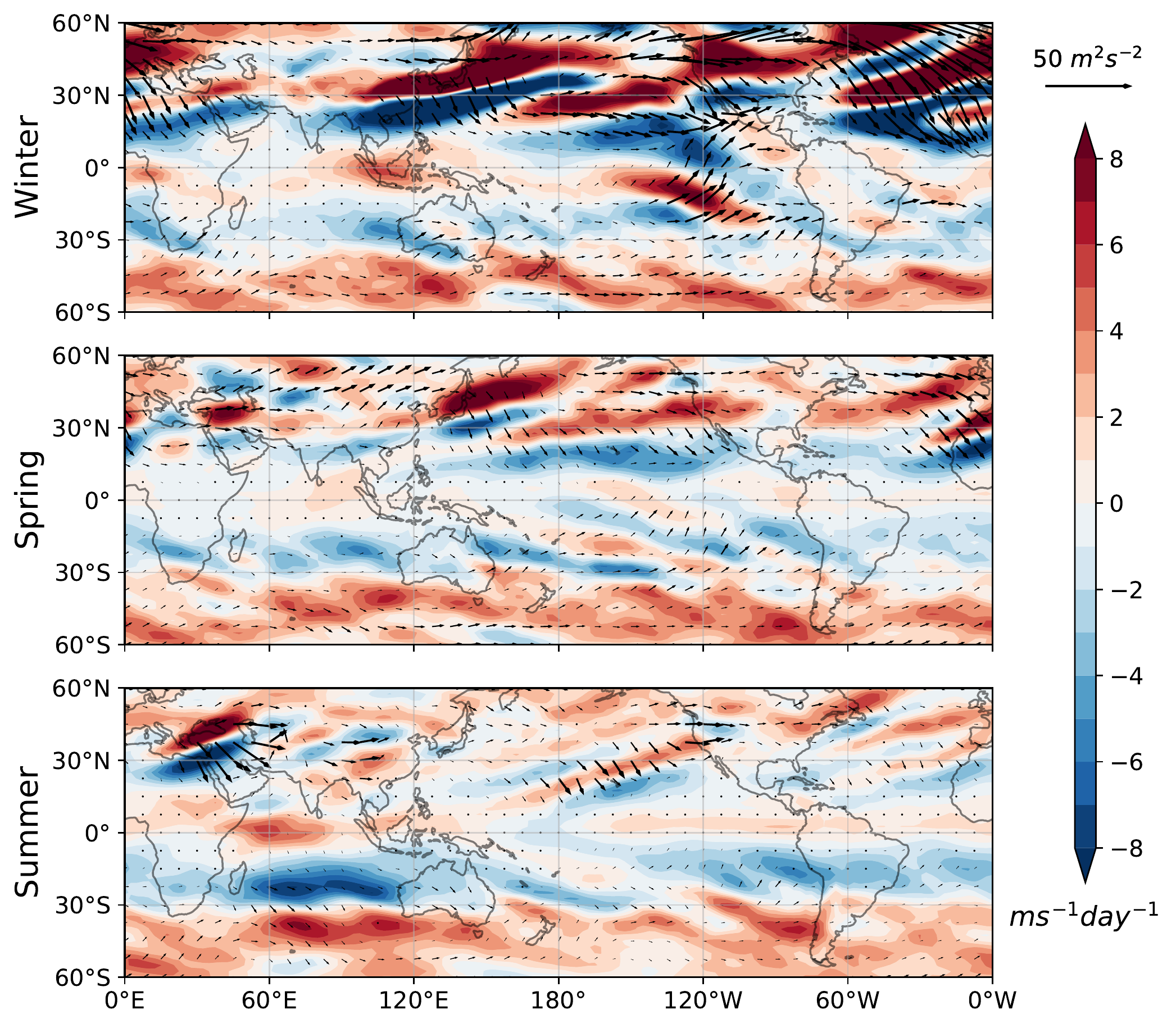}
    \caption{Wave Activity Flux (arrows) along with eddy momentum flux convergence (colors) for the individual seasons. The wave activity is computed as in Caballero and Anderson (2009).}
\end{figure}